\documentclass[aps, prd, amsmath, floats, floatfix, twocolumn,superscriptaddress, nofootinbib, showpacs,eqsecnum]{revtex4}

\usepackage{graphicx}
\usepackage{amsmath,amssymb}
\usepackage{amsfonts}
\usepackage{xspace} 
\usepackage[usenames]{color}
\usepackage{dcolumn}
\usepackage{bm}

\usepackage{ulem}
\newcommand{\Maryland}{\affiliation{Maryland Center for Fundamental
    Physics, Department of Physics, University of Maryland, College
    Park, MD 20742}}

\newcommand{\eq}{\begin{equation}}
\newcommand{\eeq}{\end{equation}}
\newcommand{\be}{\begin{equation}}
\newcommand{\ee}{\end{equation}}
\newcommand{\bea}{\begin{eqnarray}}

\newcommand{\eea}{\end{eqnarray}}
\newcommand{\bes}{\begin{subequations}}
\newcommand{\ees}{\end{subequations}}

\begin{document}

\title{Hamiltonian of a spinning test-particle in curved spacetime}

\author{Enrico Barausse} \Maryland %
\author{Etienne Racine} \Maryland %
\author{Alessandra Buonanno} \Maryland %

\begin{abstract} 
   Using a Legendre transformation, we compute the unconstrained
  Hamiltonian of a spinning test-particle in a curved spacetime at
  linear order in the particle spin. The equations of
  motion of this unconstrained Hamiltonian coincide with the
  Mathisson-Papapetrou-Pirani equations. We then use the
  formalism of Dirac brackets to derive the constrained
  Hamiltonian and the corresponding phase-space algebra in the Newton-Wigner  
  spin supplementary condition (SSC), suitably generalized to 
  curved spacetime, and find that the phase-space algebra $(\mathbf{q},\mathbf{p},\mathbf{S})$ 
  is canonical at linear order in the particle spin. We provide explicit expressions for this Hamiltonian in 
  a spherically symmetric spacetime, both in isotropic and spherical coordinates, and in the Kerr spacetime in
  Boyer-Lindquist coordinates. Furthermore, we find 
  that our Hamiltonian, when expanded in Post-Newtonian (PN) orders, agrees with the Arnowitt-Deser-Misner (ADM)
  canonical Hamiltonian computed in PN theory in the test-particle
  limit. Notably, we recover the known spin-orbit couplings through 2.5PN
  order and the spin-spin couplings of type $S_{\rm Kerr}\,S$ (and $S_{\rm Kerr}^2$) 
  through 3PN order, $S_{\rm Kerr}$ being the spin of the Kerr spacetime.  
  Our method allows one to compute the PN Hamiltonian at any order, in the test-particle limit 
and at linear order in the particle spin. As an application we compute it at 3.5PN order.
\end{abstract}

\date{\today \hspace{0.2truecm}}

\pacs{04.25.D-, 04.25.dg, 04.25.Nx, 04.30.-w}

\maketitle

\section{Introduction}
\label{sec:intro}

The dynamics of spinning bodies in general relativity is a complicated 
problem which has been investigated in several papers during the last seventy 
years, starting from the pioneering work by Mathisson~\cite{Math}, 
Papapetrou~\cite{Papa51, Papa51spin, CPapa51spin}, Pirani~\cite{Pirani}, 
Tulczyjew~\cite{Tul1,Tul2} and Dixon~\cite{Dixon}. Spin effects on the 
free motion of a test particle were first derived in the form of a 
coupling to the spacetime curvature in Refs.~\cite{Papa51, Papa51spin, CPapa51spin}. 
The computation assumes that the test-particle can be described by a 
pole-dipole energy-momentum tensor~\cite{Tul1,Tul2}, thus  
neglecting the quadrupole moment (and higher multipole moments)
and providing spin couplings only at 
linear order in the test-particle's spin.

The two-body dynamics of spinning objects can also be computed in 
post-Newtonian (PN) theory~\cite{BLiving}, which is 
basically an expansion in powers of $v/c$ and $GM/(c^2 r)$, 
where $v$ is the characteristic velocity of the system and $r$ 
is the binary's separation. Currently, spin couplings have been  computed in the two-body equations of motion 
through 2.5PN order~\cite{BOC75,BOC79,KWWi93,WWi96,TOO01,FBBu06}, and in the Arnowitt-Deser-Misner (ADM) 
canonical Hamiltonian through 3PN order~\cite{Damour-Schafer:1988,Damour:2007nc,SHS07,SSH08,SHS08} and partially 
at higher PN orders \cite{Hergt:2007ha,hergt_schafer_08}. These coupling terms agree with those  computed via effective-field-theory 
techniques at 1.5PN, 2PN and 3PN order~\cite{PR06,PR07,PR08b,PR08a}.

The main motivation for describing as accurately as possible 
the dynamics of a binary system of spinning compact bodies in general relativity comes from the forthcoming 
observation of gravitational waves with ground and space-based detectors. In particular, LIGO, Virgo 
and GEO could observe signals emitted by stellar-mass black-hole and neutron-star 
binaries, and LISA could detect signals from supermassive black-hole binaries and extreme-mass ratio binaries. 

In this paper we compute the Hamiltonian of a test-particle in a curved 
background spacetime, including all couplings linear in the test-particle's spin. Starting from the 
Lagrangian given in Ref.~\cite{porto}, we apply a Legendre transformation to derive the {\it unconstrained} 
Hamiltonian. The Hamiltonian is unconstrained in the sense that the test-particle's spin variables are  
given by an antisymmetric tensor $S^{\mu\nu}$, which a priori contains six degrees of freedom instead of three. It is well-known that in order to fix the unphysical degrees of freedom associated
with the arbitrariness in the definition of $ S^{\mu\nu}$, a choice must 
be made for the so-called spin supplementary condition (SSC). 
The arbitrariness can be interpreted, in the case of extended bodies\footnote{It should be stressed that \textit{any} spinning ``particle'' must actually have a small non-finite size. An intuitive argument for this can be found in Ref.~\cite{MTW}, Ex. 5.6, where it is shown that any spinning body must have a minimal size in order not to rotate at velocities larger than $c$. A more rigorous proof can be found in Ref.~\cite{semerak}, Sec. 2.}, as the
freedom of choosing the point, internal to the body, whose motion is followed~\cite{semerak}.

Building on the work by Hanson and Regge~\cite{hanson} and 
generalizing the Newton-Wigner (NW) SSC to curved spacetime, we then derive the {\it constrained} 
Hamiltonian and the corresponding Dirac brackets, which should replace the Poisson brackets 
when computing the equations of motion from that Hamiltonian. Quite interestingly, we find 
that the NW SSC leads, at least at linear order in the particle spin, to canonical Dirac brackets, \textit{i.e.} the standard sympletic structure for a set
of dynamical variables $(\bm{q},\bm{p},\bm{S})$. As a consistency check of 
our results we also compare our constrained Hamiltonian with the ADM canonical Hamiltonian for spinning bodies, as computed in PN theory 
through 3PN order. In addition we provide explicit expressions for the Hamiltonian of a spinning particle moving in a generic spherically symmetric spacetime (using both isotropic and spherical coordinates), as well as in the Kerr spacetime (in Boyer-Lindquist coordinates).

Another important application of this work will be developed in a subsequent paper, 
where we will use the Hamiltonian derived here to build a new effective-one-body 
Hamiltonian~\cite{Buonanno99,DJS3PN,Damour01c,DJS08} for spinning objects. This application 
is crucial to take full advantage of the analytical and numerical 
treatment of the dynamics of spinning bodies throughout the inspiral, merger and 
ringdown, and build accurate templates for the search of gravitational waves with 
ground-based and space-based detectors.

The paper is organized as follows. In Sec.~\ref{sec:notations} 
we briefly summarize our notations.  
In Sec.~\ref{sec:unconstr_hamiltonian} we apply a Legendre transformation 
to compute the unconstrained Hamiltonian and show that the equations 
of motion that follow from it coincide with the well-known Mathisson-Papapetrou-Pirani (MPP) equations 
of motion. In Sec.~\ref{sec:constr_hamiltonian}, after reviewing the 
Dirac bracket formalism, we derive the constrained Hamiltonian and the corresponding 
Dirac brackets using the generalized NW SSC. In Sec.~\ref{sec:explicit}, we
specialize our results to spherically symmetric spacetimes and to the Kerr spacetime in
Boyer-Lindquist coordinates. In Sec.~\ref{sec:hamiltonianPN} we restrict ourselves to the Kerr spacetime
in ADM coordinates, expand the Hamiltonian computed in the NW SSC in a PN series through 3.5PN order and 
find agreement with the ADM canonical Hamiltonian in the test-particle 
limit through 3PN order. Section~\ref{sec:conclusions} summarizes
our main conclusions. 

\section{Notations}
\label{sec:notations}

Throughout this paper, we will use the signature $(-,+,+,+)$ for the
metric. Spacetime tensor indices (ranging from 0 to 3) will be denoted
with Greek letters, while spatial tensor indices (ranging from 1 to 3)
will be denoted with lowercase Latin letters. Also, we will often use $t$ as alternate for the timelike index $0$.

We define a tetrad field as a set consisting of a timelike future-oriented vector $\tilde{e}^\mu_{ T}$  and three spacelike vectors 
$\tilde{e}^\mu_{ I}$ (${ I}=1,...,3$) --- 
collectively denoted as $\tilde{e}^\mu_{ A}$ (${ A}=0,...,3$) --- satisfying\footnote{We use the notation $\tilde{e}^\mu_A$ to denote any choice of tetrad given a background spacetime. The tetrad without the tilde $e^\mu_A$ refers to a special tetrad, namely the one carried by the test particle. The tetrad $e^\mu_A$ is special in the sense that it is a dynamical variable whose evolution along the worldline is prescribed by some Lagrangian.}
\begin{equation}\label{tetrad_orthonormal}
\tilde{e}^\mu_{ A}\,\tilde{e}^\nu_{ B}\, g_{\mu\nu} = \eta_{ AB}\,,
\end{equation}
where $\eta_{ TT}=-1$, $\eta_{ TI}=0$, $\eta_{ IJ}=\delta_{ IJ}$
($\delta_{ IJ}$ being the Kronecker symbol). Thus the internal
tetrad space is Lorentz invariant, \textit{i.e.} one can obtain any tetrad from an existing
one by applying a Lorentz transformation $\tilde{e}_A^\prime = \Lambda_A^{\phantom{A} B} \tilde{e}_B$, where
\begin{equation}\label{lorentz}
\Lambda_{ A}^{\phantom{A} C}\,\Lambda^{\phantom{C} B}_{C}=
\Lambda^{ C}_{\phantom{C} A}\,\Lambda_{\phantom{B} C}^{B}
=\delta^{ B}_{ A}\,.
\end{equation}
Internal tetrad indices denoted with the uppercase Latin letters ${ A}$,
${ B}$, ${ C}$ and ${ D}$ always run from $0$ to $3$, while internal tetrad indices
with the uppercase Latin letters ${ I}$, ${ J}$, ${ K}$ and ${ L}$, associated
with the spacelike tetrad vectors, run from $1$ to $3$ only. The timelike 
tetrad index is denoted by $T$.

Tetrad indices are raised and lowered with the metric $\eta_{ AB}$
[\textit{e.g.}, $\tilde{e}^\mu_{ A}=\eta_{ AB}\,(\tilde{e}^{ B})^\mu$]. With this
convention the relation~\eqref{tetrad_orthonormal} can be easily shown
to be equivalent to the completeness relation
\begin{equation}\label{tetrad_completeness}
\tilde{e}^\mu_{ A} \tilde{e}_\nu^{ A}  = \delta^\mu_\nu\,.
\end{equation}
We will denote the projections of a vector $\boldsymbol{V}$ onto the
tetrad with $V^{ A}\equiv V^\mu\, \tilde{e}^{ A}_{\mu}$, and similarly for tensors
of higher rank, as well as Christoffel symbols. Partial derivatives  will be denoted with a comma or with
$\partial$, covariant derivatives with a semicolon, 
while total covariant derivatives with respect to a parameter $\sigma$ will be denoted by $D/D\sigma$.
Finally, we will denote the operation of antisymmetrization 
with respect to the indices $\mu$ and $\nu$ as $A^{\dots [\mu}\,B^{\nu] \dots}\equiv (A^{\dots \mu}\,B^{\nu \dots}-A^{\dots \mu}\,B^{\nu \dots})/2$.

We use geometric units $G=c=1$ throughout the paper, except in Sec.~\ref{sec:hamiltonianPN} 
where the factors of $c$ are restored, playing the role of PN book-keeping parameters.

\section{Unconstrained Hamiltonian}
\label{sec:unconstr_hamiltonian}

In this section we derive the unconstrained Hamiltonian by applying a Legendre 
transformation to the Lagrangian describing the motion of a spinning particle 
in a generic curved spacetime.

\subsection{The Lagrangian and the Mathisson-Papapetrou-Pirani  equations}

Building on the classic work of Hanson and Regge~\cite{hanson} which analyzes the dynamics of a relativistic top in a flat spacetime, 
Porto showed in Ref.~\cite{porto} that the equations of motion of a spinning particle in curved spacetime can be
obtained from the action 
\begin{equation}
\label{portoaction}
S=\int L(a_1,a_2,a_3,a_4)\, d\sigma\,,
\end{equation}
$\sigma$ being a parameter along the representative worldline. The Langrangian $L$ is a function of the four Lorentz-invariant scalars
\begin{eqnarray}
a_1&=&u_\mu\, u^\mu\,,\\
a_2&=&\Omega_{\mu\nu}\, \Omega^{\mu\nu}\,,\\
a_3&=&u_\mu\,\Omega^{\mu\nu}\, \Omega_{\nu\rho}\, u^\rho\,,\\
a_4&=&\Omega_{\lambda\mu} \, \Omega^{\mu\nu}\, \Omega_{\nu\rho}\, \Omega^{\rho\lambda}\,,
\end{eqnarray}
where $u^\mu\equiv d x^\mu/d\sigma$ is the tangent vector to the representative worldline, 
and where the antisymmetric tensor $\Omega^{\mu\nu}$ describes how the tetrad $e^\mu_A$ carried by the particle rotates along the worldline:
\begin{equation}\label{omega_def}
\Omega^{\mu\nu}=\eta^{ AB}\, e_{ A}^{\mu}\, \frac{D e_{ B}^{\nu}}{D\sigma}=
{e}^{ A\mu}\,\frac{d{e}_{ A}^{\nu}}{d\sigma} +\Gamma^\mu_{\alpha\beta}\,g^{\alpha\nu}\,u^\beta\,.
\end{equation} 
Moreover, the action~\eqref{portoaction} is assumed to be
reparametrization-invariant (\textit{i.e.}~its form must be
independent of the particular parameter used to follow the particle's
worldline), which translates in the requirement that the Lagrangian
$L$ be a homogeneous function of degree one in the ``velocities''
$u^\mu$ and $\Omega^{\mu\nu}$~\cite{hanson}. Porto then
shows that if one \textit{defines} the four-momentum vector and the
spin tensor of the particle as\footnote{Because of
    reparametrization invariance of the action~\eqref{portoaction},
    these definitions maintain the same form whatever parameter
    $\sigma$ is chosen along the wordline, as appropriate for physical
    quantities like the four-momentum and the spin.} 
\begin{eqnarray}
&& p_\mu\equiv \frac{\partial L}{\partial u^\mu}\Big\vert_\Omega\,,\label{pdef}\\
&& S_{\mu\nu}\equiv 2 \frac{\partial L}{\partial
  \Omega^{\mu\nu}}\Big\vert_u\label{sdef} 
\end{eqnarray} 
[note that $p_\mu$ is \textit{not} the momentum conjugate to the coordinates $x^\mu$ because
$\Omega^{\mu\nu}$ depends on $u^\mu$, as can be seen in
Eq.~\eqref{omega_def}], then a variation of the action with respect to $e_A^{\mu}$ which preserves 
the defining property~\eqref{tetrad_orthonormal} of a tetrad gives the precession equation
for the spin tensor 
\begin{equation}
\label{spinevolution} \frac{D S^{\mu\nu}}{D\sigma}=
S^{\mu\lambda}\,\Omega_\lambda^{\nu}-\Omega^{\mu\lambda}\,S_\lambda^{\nu}=p^\mu\,u^\nu-p^\nu\,u^\mu\,. 
\end{equation} 
The second equality in Eq.~\eqref{spinevolution} follows from definitions~\eqref{pdef} and~\eqref{sdef}, 
and from the fact that the Lagrangian in the action~\eqref{portoaction} depends only on
$a_1$, $a_2$, $a_3$ and $a_4$~\cite{hanson}. Moreover, a variation of the action
with respect to the particle's position $x^\mu$ gives~\cite{porto}
\begin{equation}\label{pevolution} \frac{D p^{\mu}}{D\sigma}=-\frac12
R^\mu_{\phantom{\mu}\alpha\beta\gamma}\,u^\alpha\,S^{\beta\gamma}\,.
\end{equation} 
Thus one precisely recovers the
well-known MPP equations from the action~\eqref{portoaction}, which therefore encodes the
dynamics of a spinning test-particle in  curved spacetime, at linear order in the particle's spin.

Notice however that the set of Eqs.~\eqref{spinevolution}--\eqref{pevolution} 
consists of ten equations and thirteen independent variables ($p^\mu$, $u^\mu$ and $S^{\mu\nu}$, 
subject to the normalization constraint\footnote{For example one is free to select a parameter 
$\sigma = \tau$ such that $u_{\mu} u^{\mu} = -1$, since the action~\eqref{portoaction} 
is reparametrization invariant. Any other choice of parameter simply yields a 
different normalization constraint $u_{\mu}u^{\mu} = -(d\sigma/d\tau)^{-2}$.}
of the tangent vector $u^\mu$) and is
therefore not closed. This underdetermination can be addressed by
imposing a SSC, which is typically expressed as 
\begin{equation}\label{SSCstandard}
S^{\mu\nu}\omega_{\nu} = 0, 
\end{equation}
where $\omega_\nu$ is some suitably chosen timelike 
vector. Equation~\eqref{SSCstandard} contains three independent constraints,
and is therefore expected to reduce the number of independent variables from 13 to 10, thus closing the system 
of Eqs.~\eqref{spinevolution}--\eqref{pevolution}. This is indeed what happens, as
the requirement that Eq.~\eqref{SSCstandard} be valid at all points along
the worldline implies the following implicit relationship between $p^\mu$ and $u^\mu$
\begin{equation}\label{pu_relation}
p^\mu = \frac{1}{\omega_\nu u^\nu}\left[(\omega_\nu p^\nu)u^\mu - S^{\mu\nu}\frac{D\omega_\nu}{D \sigma} \right]\,.
\end{equation}

It should be stressed once again that it is the \textit{underdetermination} 
of the unconstrained MPP system that allows one to impose \textit{any} constraint of the 
form~\eqref{SSCstandard}, 
and that the constraint will be \textit{automatically} conserved by the time evolution of the system 
because of Eq.~\eqref{pu_relation}. Of course 
 different constraints of the form~\eqref{SSCstandard}
will produce different systems of equations describing the evolution of the particle's worldline. The physical reason for this is easy to understand: the SSC~\eqref{SSCstandard}  binds the test-particle described by the  Lagrangian 
to a specific, SSC-dependent, worldline lying inside the worldtube spanned by the spinning body, namely the center of energy of the body as seen by an observer with
four-velocity parallel to $\omega^\mu$ (see \textit{e.g.} Ref.~\cite{semerak}
for a lucid discussion of the physical meaning of SSCs).

\subsection{Deriving the Hamiltonian through a Legendre transformation}

It is convenient to rewrite the action~\eqref{portoaction} as 
\begin{equation}\label{action}
S=\int L\left(x^\mu,u^\mu,\phi^a, \frac{d\phi^a}{d\sigma}\right)\, d\sigma\,,
\end{equation}
where the Langrangian $L$ can be now considered as a function 
of the coordinates $x^\mu$, the four-vector $u^\mu= d x^\mu/d\sigma$, the six parameters $\phi^a$ and 
their time derivatives. The set $\{\phi^a\}$ consists simply
of the parameters of the internal Lorentz transformation
describing the orientation  of the tetrad field $e_{ A}^\mu$ carried by the
particle with respect to an arbitrary, but fixed,
reference tetrad field $\tilde{e}_{ A}^\mu(x)$ covering the whole
spacetime~\footnote{In what follows we will prove that the equations of motion are independent of the choice 
  of this tetrad field. This had to be expected, based on Refs.~\cite{porto,hanson}.}. Therefore, the
tetrad carried by the particle is given by
\begin{equation}
e_{ A}^\mu(\phi,x)=\Lambda_{ A}^{\phantom{A} B}(\phi)\,\tilde{e}_{ B}^\mu(x)\,,
\end{equation}
where $\Lambda_{ A}^{\phantom{A} B}$ is a Lorentz transformation.
We also note that the parameters
$\phi^a$ and their time derivatives enter the Lagrangian only through
the antisymmetric tensor $\Omega^{\mu\nu}$, which we write explicitly
as
\begin{eqnarray}
 \label{omega_long}
\Omega^{\mu\nu} &=& \eta^{AB}\,e^\mu_A(\phi,x)\,\left [\frac{d \phi^a}{d \sigma}\,\frac{\partial e^\nu_{B}}{\partial\phi^a}(\phi,x) + u^\beta\,e^{\nu}_{B,\beta}(\phi,x) \right ] \nonumber \\
&& + \Gamma^\nu_{\alpha\beta}\,g^{\mu\alpha}\,u^\beta\,.
\end{eqnarray}
To construct the Hamiltonian we need to choose a particular 3+1 decomposition of the background metric. We take $\sigma=t$, where $t$ is the time coordinate of that particular decomposition. 
Using reparametrization invariance, we can write
\begin{eqnarray}\label{rep_inv}
S&=&\int L\left(x^\mu,u^\mu,\phi^a, \frac{d\phi^a}{d\sigma}\right)\, d\sigma\,, \nonumber \\
&=& \int L(x^i,v^i,\phi^a,\dot{\phi}^a,t)\,dt\,,
\end{eqnarray}
where $x^0 = t$, $u^0 = 1$, $u^i=v^i = {dx^i}/{dt}$ and $\dot{\phi}^a = {d\phi^a}/{dt}$.
The configuration space of the spinning particle therefore consists of the set $\{x^i,\phi^a\}$. 
The total variation of the Lagrangian considered as function of $x^i,v^i, \phi^a$ and $\dot{\phi}^a$ is
\begin{eqnarray}
\label{deltaL1}
\delta L&=& \frac{\partial L}{\partial x^i}\,\delta x^i + 
\frac{\partial L}{\partial v^i}\,\delta v^i + 
\frac{\partial L}{\partial \phi^a}\,\delta \phi^a + 
\frac{\partial L}{\partial \dot{\phi}^a}\,\delta \dot{\phi}^a\,, \nonumber \\
&\equiv&  \frac{\partial L}{\partial x^i}\,\delta x^i + P_i\,\delta v^i + 
\frac{\partial L}{\partial \phi^a}\,\delta \phi^a + 
P_{\phi^a}\,\delta \dot{\phi}^a \,,
\end{eqnarray}
where we denoted by $P_i$ and $P_{\phi^a}$ the momenta conjugate to $x^i$ and $\phi^a$, respectively.
The total variation of the Lagrangian considered as function of $x^i, v^i$ and $\Omega^{\mu \nu}$ is instead
\begin{equation}
\label{deltaL2}
\delta L= \frac{\partial L}{\partial x^i}\Big \vert_\Omega\,\delta x^i +  
\frac{\partial L}{\partial v^i}\Big \vert_\Omega\,\delta v^i + 
\frac{\partial L}{\partial \Omega^{\mu \nu}}\Big \vert_{x,v}\,\delta \Omega^{\mu \nu}\,. 
\end{equation}
Using Eq.~\eqref{omega_long}, Eq.~\eqref{deltaL2} can be rewritten as 
\begin{eqnarray}
\label{deltaL3}
\delta L&=& \left (\frac{\partial L}{\partial x^i}\Big \vert_\Omega + 
\frac{\partial L}{\partial \Omega^{\mu \nu}}\Big \vert_{x,v}\,\frac{\partial \Omega^{\mu \nu}}{\partial x^i}\Big \vert_{v,\phi,\dot{\phi}} 
\right ) \,\delta x^i \nonumber \\  
&+& \left (\frac{\partial L}{\partial v^i}\Big \vert_\Omega + 
\frac{\partial L}{\partial \Omega^{\mu \nu}}\Big \vert_{x,v}\,\frac{\partial \Omega^{\mu \nu}}{\partial v^i}\Big \vert_{x,\phi,\dot{\phi}} 
\right ) \,\delta v^i \nonumber \\
&+& \frac{\partial L}{\partial \Omega^{\mu \nu}}\Big \vert_{x,v}\,\frac{\partial \Omega^{\mu \nu}}{\partial \phi^a}\Big \vert_{x,v,\dot{\phi}} \,\delta \phi^a \nonumber \\
&+& \frac{\partial L}{\partial \Omega^{\mu \nu}}\Big \vert_{x,v}\,\frac{\partial \Omega^{\mu \nu}}{\partial \dot{\phi}^a}\Big \vert_{x,v,{\phi}} \,\delta \dot{\phi}^a\,. 
\end{eqnarray}

Comparing Eq.~\eqref{deltaL1} with Eq.~\eqref{deltaL3}, and using Eqs.~\eqref{pdef}, \eqref{sdef} 
and Eq.~\eqref{omega_long}, we obtain the conjugate momenta
\begin{eqnarray}
P_i &=& p_i + \frac{1}{2}\,\eta^{AB}\,S_{\mu \nu}\,e^\mu_A\,e^\nu_{B;i}\,,\nonumber \\
 &=& p_i + \frac{1}{2}\,\eta^{AB}\,S_{\mu \nu}\,\tilde{e}^\mu_A\,\tilde{e}^\nu_{B;i}\,, \nonumber \\
&\equiv& p_i + E_{i\mu\nu} S^{\mu\nu}\,, \label{Pic}
\end{eqnarray}
and
\begin{eqnarray}
P_{\phi^a} &=& \frac{1}{2}\,\eta^{AB}\,S_{\mu \nu}\,e^\mu_A\,\frac{\partial e^\nu_{B}}{\partial \phi^a}\,,\nonumber \\
&=& \frac{1}{2}\,S_{\mu \nu}\,\lambda_a^{AB}\,\tilde{e}^\mu_A\,\tilde{e}^\nu_B\,,
\label{Pphic}
\end{eqnarray}
where we have introduced the tensor
\begin{equation}\label{Edef}
E_{\lambda\mu\nu}\equiv  \frac{1}{2}\,\eta_{AB}\,\tilde{e}_\mu^A\,\tilde{e}_{\nu;\lambda}^{B}\,,
\end{equation}
which is antisymmetric in the last two indices, and the antisymmetric tensor~\cite{hanson}
\begin{equation}\label{lambda_def}
\lambda_a^{ AB}(\phi)\equiv
\Lambda_{C}^{\phantom{A} A}\,\frac{\partial \Lambda^{ CB}}{\partial \phi^a}\,.
\end{equation}

A necessary condition to go from the Lagrangian formalism to the Hamiltonian one in the usual way (\textit{i.e.} by means of a 
Legendre transformation) is that the Langrangian is \textit{regular}~\cite{goldstein}, \textit{i.e.} it satisfies\footnote{
While this condition is sufficiently generic to leave our Lagrangian essentially undetermined, it should be noticed that there 
are famous examples in physics where this regularity condition does not hold, such as the electromagnetic field 
(see for instance Ref.~\cite{sundermeyer}, chapter 5), the Dirac field~(see for instance Ref.~\cite{peskin}, problem 9.2d), 
the Schrodinger equation (see Ref.~\cite{gergely} and references therein) and general relativity 
(see for instance Ref.~\cite{sundermeyer}, chapter 9).
}
\eq
\label{det}
{\rm det} \left(\frac{\partial^2 L}{\partial \dot q^i\partial \dot q^j}\right)\neq0\,,
\eeq
where $\mathbf{q}=(x^i,\phi^a)$. Under this condition, we can perform the usual Legendre transformation to get the Hamiltonian
\begin{equation}\label{Hdef}
H=P_i\,v^i + P_{\phi^a}\,\dot{\phi}^a - L\,. 
\end{equation}
Since $L$ is homogeneous of degree one in the ``velocities'' [because of the reparametrization invariance of the action~\eqref{portoaction}], 
Euler's theorem  implies that
\begin{eqnarray}
&&u^\mu\,\frac{\partial L}{\partial u^\mu}\Big \vert_\Omega + 
\Omega^{\mu \nu}\, \frac{\partial L}{\partial \Omega^{\mu\nu}}\Big \vert_u \nonumber \\
&&=v^i\,p_i+ p_t + \frac{1}{2} \,S_{\mu \nu}\,\Omega^{\mu \nu} = L\,,
\label{euler}
\end{eqnarray}
where we have used the definitions~\eqref{pdef} and~\eqref{sdef}, as well as the fact that with our time-slicing $u^0=1$ and $u^i=v^i$.
Using now Eqs.~\eqref{euler},~\eqref{Pic},~\eqref{Pphic} and~\eqref{omega_long} (with $u^0=1$ and  $u^i=v^i$) in Eq.~\eqref{Hdef},
simple algebra allows one to write the Hamiltonian as
\begin{subequations}
\begin{eqnarray}
H&=&-p_t-\frac12 \eta^{AB} S_{\alpha\beta}\,{e}_{ A}^\alpha\,{e}^{\beta}_{B;t}\,, \label{HamiltonianI}\\
&=&-p_t-\frac12 \eta^{AB} S_{\alpha\beta}\,\tilde{e}_{ A}^\alpha\,\tilde{e}^{ \beta}_{B;t}\,, \label{HamiltonianII}\\
&\equiv& - p_t - E_{t\mu\nu}S^{\mu\nu}\,,\label{Hamiltonian} 
\end{eqnarray}
\end{subequations}
where the covariant derivative with respect to $t$ in the second term 
of Eq.~\eqref{HamiltonianI} above is a shorthand for covariant derivative 
with respect to $x^0=t$, \textit{i.e.} one can pull the Lorentz transformation
$\Lambda_A^{\phantom{A} B}(\phi)$ outside the covariant derivative
as it is independent of $x^0$. 
It should be noted that using the tensor $E_{\lambda\mu\nu}$ defined in Eq.~\eqref{Edef}, one can combine 
$H$ and $P_i$ into a four-vector $P_\alpha$ such that
\begin{equation}\label{capP4v}
P_\alpha = (-H,P_i) = p_\alpha + E_{\alpha\mu\nu} S^{\mu\nu}\,.
\end{equation}

The MPP equations of motion can be derived 
from the Hamiltonian \eqref{Hamiltonian} as follows. On one hand we have 
\begin{equation}
\frac{d P_{\phi^a}}{dt} = -\frac{\partial H}{\partial \phi^a} = 
\frac{\partial L}{\partial \phi^a}\Big \vert_{x,u,\dot{\phi}} = 
\frac{\partial L}{\partial \Omega^{\mu \nu}}\Big \vert_{x,u}\, 
\frac{\partial \Omega^{\mu \nu}}{\partial \phi^a}  
\,,
\end{equation}
where the second equality follows from the definition of the Hamiltonian~\eqref{Hdef} with the regularity condition~\eqref{det}. 
(One could also derive the second equality by comparing the Hamiltonian and Lagrange equations, but it should be stressed that these two sets of equations are equivalent only if
the  regularity condition~\eqref{det} is satisfied~\cite{goldstein}.)
Using then Eqs.~\eqref{omega_long} and~\eqref{Pphic}, as well as the definition~\eqref{sdef}, a straightforward 
computation gives the precession equation
\begin{equation}
\frac{D S^{\mu\nu}}{Dt}=S^{\lambda\mu}\,\Omega_{\,\,\,\lambda}^{\nu}-\Omega^{\mu\lambda}\,S_{\,\,\,\lambda}^{\nu}\,.
\end{equation} 
The translational equations of motion can be obtained following a similar procedure. 
In the neighborhood of any event located on the particle's worldline we can choose Riemann normal coordinates and write
\begin{eqnarray}
\frac{d P_{i}}{dt} &=& \frac{d p_{i}}{dt} + \frac{1}{2} \frac{d}{dt} \left ( S_{\mu \nu}\,\eta^{AB}\,
e_A^\mu\,e^\nu_{B;i} \right )\,,\nonumber \\
&=& -\frac{\partial H}{\partial x^i} = 
\frac{\partial L}{\partial x^i}\Big \vert_{u,\phi,\dot{\phi}}\,, \nonumber \\
&=& 
\frac{\partial L}{\partial \Omega^{\mu \nu}}\Big \vert_{x,u}\, 
\frac{\partial \Omega^{\mu \nu}}{\partial x^i} \,,
\end{eqnarray} 
where the last equality follows from the compatibility of the metric with the connection, i.e. $g_{\mu\nu;i} = 0$, which becomes
$g_{\mu\nu,i} = 0$ in Riemann normal coordinates.\footnote{We stress that one is allowed to set $g_{\mu\nu,i} = 0$ in this equation as 
we do not need to take derivatives of it (in which case, of course, the terms containing  $g_{\mu\nu,i}$ would give a contribution, 
as in general $g_{\mu\nu,ij} \neq 0$ even in  Riemann normal coordinates).} Making use of 
Eq.~\eqref{omega_long} and using the fact that in Riemann normal coordinates $\Gamma_{\mu\nu}^\lambda=0$, 
while their derivatives are non-zero, we get
\begin{equation}\label{p_ievol_local}
\frac{d p_i}{dt}=-\frac12
R_{i\alpha\beta\gamma}\,u^\alpha\,S^{\beta\gamma}\,,
\end{equation}
where the Riemann tensor term arises from the derivatives of the Christoffel symbols 
appearing in Eq.~\eqref{omega_long}. Rewriting Eq.~\eqref{p_ievol_local} in a generic coordinates system, we immediately get the spatial part of the translational MPP equations
\begin{equation}\label{p_ievol}
\frac{D p_i}{Dt}=-\frac12
R_{i\alpha\beta\gamma}\,u^\alpha\,S^{\beta\gamma}\,.
\end{equation}
The unconstrained equation of motion for $p_t$ is obtained as follows. One starts from the formal expression
\begin{equation}
\frac{dp_t}{dt} = \{p_t,H\} + \frac{\partial p_t}{\partial t} \,.
\end{equation}
In Riemann normal coordinates, the left-hand side is equal to ${Dp_t}/{Dt}$. To evaluate the right-hand side, one makes use of Eq.~\eqref{Hamiltonian} to eliminate $p_t$ in favor of the Hamiltonian and other quantities whose explicit expressions in terms of the phase-space variables $\{x^i,P_i,\phi^a,P_{\phi^a}\}$ are known. Straightforward algebra then yields
\begin{equation}\label{p_tevol}
\frac{Dp_t}{Dt} = -\frac{1}{2}R_{t\alpha\beta\gamma}u^\alpha S^{\beta\gamma}\,,
\end{equation}
which can be combined with Eqs.~\eqref{p_ievol} in the well-known equation translational MPP equations
\begin{equation}
\frac{Dp_\mu}{Dt} = -\frac{1}{2}R_{\mu\alpha\beta\gamma}u^\alpha S^{\beta\gamma}\,.
\end{equation}
Before concluding this section, we provide explicit expressions for the Poisson brackets of the variables $x^i, P_i$, $S^{\mu \nu}$ and 
$\Lambda^{\mu\nu}\equiv \Lambda^{AB} \tilde{e}_A^\mu \tilde{e}_B^\nu$. Using the definition of Poisson bracket, 
\begin{equation}
\{f,g\}\equiv \frac{\partial f}{\partial \mathbf{q}}\cdot\frac{\partial
  g}{\partial \boldsymbol{\pi}}-\frac{\partial g}{\partial \mathbf{q}}\cdot\frac{\partial
  f}{\partial \boldsymbol{\pi}}\,, 
\end{equation} 
where $\mathbf{q}=(x^i,\phi^a)$ and $\boldsymbol{\pi}=(P_i,P_{\phi^a})$, we trivially have 
\begin{subequations}\label{xPalgebra_unconstrained}
\begin{eqnarray}\label{xp}
\{x^i,P_j\}&=&\delta^i_j\,,\\
\{x^i,x^j\}&=&\{P_i,P_j\}=0\,.\label{xx} 
\end{eqnarray} 
\end{subequations}
To compute the Poisson brackets involving $S^{\mu\nu}$, let us first
invert Eq.~\eqref{Pphic}~\cite{hanson}:
\begin{equation}\label{spin_tensor}
S^{\alpha\beta}=\tilde{e}^\alpha_{ A}\,\tilde{e}^\beta_{ B}\, \rho^{ AB}_a\,P_{\phi^a}\,, 
\end{equation} 
where $\rho_a^{AB}(\phi)$ satisfies 
\begin{eqnarray}
\rho_a^{AB}\,\lambda_{b AB}&=&2\delta_{ab}\,,\\
\lambda_a^{AB}\,\rho_a^{CD}&=&\eta^{AC}\,\eta^{BD}-\eta^{AD}\,\eta^{BC}.  
\end{eqnarray} 
Using these relations together with the identity
\begin{equation}\label{a_comm}
\frac{\partial \lambda_a^{ AB}}{\partial \phi^b}-\frac{\partial
  \lambda_b^{ AB}}{\partial
  \phi^a}=\lambda_a^{ AC}\,\lambda_{b\, C}^{\phantom{AB} B}-\lambda_b^{ AC}\,\lambda_{a\, C}^{\phantom{AB} B}\,,
\end{equation} 
which can be immediately derived~\cite{hanson} by taking the derivative of Eq.~\eqref{lambda_def},
it is straightforward to prove that $\rho_a^{AB}$ is a realization of the Lie algebra of the Lorentz group:
\begin{eqnarray} 
\rho_b^{AB}\,\frac{\partial \rho_a^{CD}}{\partial
  \phi^b}-\rho_b^{CD}\,\frac{\partial \rho_a^{AB}}{\partial \phi^b}&=& -\rho_a^{AC}
\eta^{BD} - \rho_a^{BD}\,\eta^{AC} \nonumber \\ 
&& \!\!\!\! +\rho_a^{AD}\,\eta^{BC}+\rho_a^{BC}\,\eta^{AD}. \nonumber \\
\end{eqnarray} 
Simple algebra then yields
\begin{subequations}\label{Salgebra_unconstrained}
\begin{eqnarray}\label{SS} 
\{S^{\mu\nu}, S^{\alpha\beta}\}&=&  S^{\mu\alpha}\,g^{\nu\beta}+S^{\nu\beta}\,
g^{\mu\alpha} \nonumber \\
&& -S^{\mu\beta}\, g^{\nu\alpha}-S^{\nu\alpha}\, g^{\mu\beta}\,,
\end{eqnarray} 
while using Eqs.~\eqref{tetrad_completeness} and~\eqref{spin_tensor} we easily obtain
\begin{eqnarray}\label{Sp}\{S^{\mu\nu}, P_i\}&=& S^{\alpha\nu}\, \tilde{e}_{\alpha}^{ A}\,
\tilde{e}_{A,i}^\mu+ S^{\mu\alpha}\,\tilde{e}_{\alpha}^{ A}\,\tilde{e}_{ A,i}^\nu\,,\\
\{S^{\mu\nu}, x^i\}&=&0\,.\label{Sx} 
\end{eqnarray} 
\end{subequations}

Finally, it is straightforward to show that $\Lambda^{AB}$ satisfies~\cite{hanson}
\begin{align}
&\{\Lambda^{AB},x^i\}=\{\Lambda^{AB},P_i\}=\{\Lambda^{AB},\Lambda^{CD}\}=0\,,\\
&\{\Lambda^{AB},S^{CD}\}=\Lambda^{AC}\eta^{BD}-\Lambda^{AD}\eta^{BC}\,,
\end{align}
or, in terms of $\Lambda^{\mu\nu}\equiv \Lambda^{AB} \tilde{e}_A^\mu \tilde{e}_B^\nu$
\begin{align}
&\{\Lambda^{\mu\nu},x^i\}=\{\Lambda^{\mu\nu},\Lambda^{\alpha\beta}\}=0\,,\label{lambdax}\\
&\{\Lambda^{\mu\nu},P_i\}=\Lambda^{\alpha\nu}\, \tilde{e}_{\alpha}^{ A}\,\tilde{e}_{A,i}^\mu+ \Lambda^{\mu\alpha}\,\tilde{e}_{\alpha}^{ A}\,\tilde{e}_{ A,i}^\nu \,,\label{lambdaP}\\
&\{\Lambda^{\mu\nu},S^{\alpha\beta}\}=\Lambda^{\mu\alpha}g^{\nu\beta}-\Lambda^{\mu\beta}g^{\nu\alpha}\label{lambdaS}\,.
\end{align}

\section{Constrained Hamiltonian}
\label{sec:constr_hamiltonian}

\subsection{Imposing constraints in phase-space: a Dirac-bracket primer}
\label{sec:DBprimer}

Let us briefly recall how constraints are imposed in the Hamiltonian formalism  
(a very detailed review on the subject can be found in Ref.~\cite{Henneaux:1992ig}).
Let us consider a Hamiltonian $H(\mathbf{q},\boldsymbol{\pi},t)$ living in a $2n$-dimensional phase space
 and a binary ``bracket'' operation $\{...,...\}$
which is antisymmetric, bilinear, and which satisfies the Leibniz rule, as well as the Jacobi identity, \textit{i.e.}
\begin{subequations}\label{bracket_properties}
\begin{eqnarray}
\{A, B\}&=&-\{B, A\}\,,\\
\{a A+ b B, C\}&=& a\{A, C\}+b\{B, C\}\,,\\
\{AB, C\}&=& \{A, C\}B+A\{B, C\}\,,
\end{eqnarray}
and
\begin{equation}
\{A,\{B, C\}\}+\{B,\{C, A\}\}+\{C,\{A, B\}\}=0\,.
\end{equation}
\end{subequations} 
In Eqs.~\eqref{bracket_properties}, $A$, $B$ and $C$ are arbitrary phase-space functions, while $a$ and $b$
are constants. Let us also assume that the bracket operation 
gives the equations of motion for a generic phase-space 
function $A$ through the Hamilton equations
\begin{equation}
\frac{dA}{dt}=\frac{\partial A}{\partial t}+\{A,H\}\,.
\end{equation}
If we consider now a set of constraints $\xi_i=0$, $i=1,...,2m$ (with $m<n$)
such that the matrix
\begin{equation}\label{CIJ}
C_{ij}\equiv \{\xi_i,\xi_j\}
\end{equation}
is not singular\footnote{In the literature, constraints satisfying this condition 
are known as \textit{second class constraints}~\cite{Henneaux:1992ig}.}, these constraints can be imposed simply by 
replacing the original brackets with the so-called \textit{Dirac
  brackets}. The Dirac brackets are in essence the projection of the
original symplectic structure onto the phase-space surface defined by the
constraints. For two arbitrary
phase-space functions $A$ and $B$, the Dirac brackets are given by
\begin{equation}\label{DB}
\{A,B\}_{\rm DB}=\{A,B\}+\{A,\xi_i\}\,\{B,\xi_j\}\,[C^{-1}]_{ij}\,.
\end{equation}
It can be shown (see \textit{e.g.} Secs.~1.3.2, 1.3.3, and Ex. 1.12 
in Ref.~\cite{Henneaux:1992ig}), 
that the Dirac brackets are bilinear, antisymmetric, that they satisfy the Leibniz rule and the Jacobi
identity, and that they provide the correct equations of motion for
the constrained system through the Hamilton equations
\begin{equation}
\frac{dA}{dt}=\frac{\partial A}{\partial t}+\{A,\bar{H}\}_{\rm DB}\,,
\end{equation}
where $A$ is an arbitrary phase-space function, and where the new 
Hamiltonian $\bar{H}$ is obtained simply by inserting the constraints in the original Hamiltonian $H$.

In summary, given a Hamiltonian $H$ and a bracket operation (\textit{e.g.}, 
the Poisson brackets in the case of an unconstrained Hamiltonian), in order to impose a set of constraints 
satisfying $\det(C_{ij})\neq 0$ [with $\boldsymbol{C}$ given by Eq.~\eqref{CIJ}], we need 
to replace the original bracket operation with the Dirac bracket operation~\eqref{DB}, and 
insert the constraints directly in the original Hamiltonian.

In Secs.~\ref{sec:NWDB} and~\ref{sec:hamiltonian} we start from the
unconstrained Hamiltonian~\eqref{Hamiltonian} and the
unconstrained algebra \eqref{xp}, \eqref{xx}, \eqref{SS}, \eqref{Sp},
\eqref{Sx}, \eqref{lambdax}, \eqref{lambdaP}, \eqref{lambdaS}, and use the procedure outlined in this subsection to impose
the generalized NW SSC. In particular, in
Sec.~\ref{sec:NWDB} we compute the Dirac brackets in the NW
SSC, showing that they are canonical (\textit{i.e.,} they reduce to the usual Poisson brackets) at linear order in the particle's spin, 
while in Sec.~\ref{sec:hamiltonian} we explicitly write the constrained
Hamiltonian. 

\subsection{Dirac brackets in the generalized Newton-Wigner spin supplementary condition}
\label{sec:NWDB}

In this section, we consider the NW SSC generalized to curved spacetime, 
\begin{equation}\label{eq:NW}
V^\mu\equiv S^{\mu \nu}\,\omega_\nu=0\,,
\end{equation}	
with 
\begin{equation}\label{eq:omegaNW}
\omega_\mu=p_\mu- m\,\tilde{e}^{T}_\mu\,,
\end{equation}
where $m = \sqrt{-p_\mu p^\mu}$ is a function of phase space variables
that we define as the mass of the particle\footnote{Note that at this
  stage there is no guarantee that this function on phase space is a
  constant of motion. We will show later that it is indeed the case,
  but we emphasize that this is a non-trivial result.}.  We
  stress that the vector $\boldsymbol{\omega}$ is the sum of two
  timelike future-oriented vectors and is therefore timelike itself,
  which implies that Eqs.~\eqref{eq:NW} and~\eqref{eq:omegaNW} do
  indeed yield a legitimate SSC~\cite{semerak}. (We recall that with
  our notation one has
  $\boldsymbol{\tilde{e}}^T=-\boldsymbol{\tilde{e}}_T$, and that
  $\boldsymbol{\tilde{e}}_T$ is future oriented.)

While the NW SSC is well-known to be the \textit{only} SSC condition which yields canonical variables in flat
spacetime\footnote{We note that in quantum mechanics and flat spacetime the NW 
SSC holds a special place~\cite{NW_paper,Pryce:1948pf}.}
~\cite{hanson,Pryce:1948pf,NW_paper}, there is no a priori
guarantee that this is the case in curved spacetime.
In this section we show that the NW SSC \textit{does} indeed yield canonical variables at linear order in the particle's spin. 

Because $V^\mu\,\omega_\mu=0$, only three of the four constraints~\eqref{eq:NW} are independent.
Since $\boldsymbol{\omega}$ is a timelike vector, it is 
natural to take the three independent constraints to be the spatial
components $V^i$. The constraints $V^i$ may be viewed as constraints on the momenta $P_{\phi^a}$, as there is a one-to-one
mapping between the spin tensor $S^{\mu\nu}$ and the six momenta conjugate to the $\phi^a$'s.
This implies that by themselves, the constraints $V^i$ do not form a consistent set of
constraints on phase-space: an additional set of three constraints must be imposed
on the configuration coordinates $\phi^a$ themselves in order to retain a symplectic structure, \textit{i.e.} that the
constraint hypersurface contains the same number of configuration coordinates and conjugate momenta. 
The additional constraints we choose to impose are given by~\cite{hanson,PR06}
\begin{eqnarray}
\chi_\mu&=&(e_{T})_{\mu}-\frac{p_\mu}{m} \nonumber \\
&=& \Lambda_{T}^{\phantom{T} A}\,(\tilde{e}_{A})_{\mu}-\frac{p_\mu}{m}=0\,. \label{constraint:chi}
\end{eqnarray}
It is worth pointing out once again that the mass $m$ is a function on phase space, and therefore its Poisson brackets with coordinates and momenta are non-vanishing. It will acquire a special status as a constant of motion (at linear order in spin) only at the end of this subsection. Equation~\eqref{constraint:chi} may be alternatively rewritten as
\begin{equation}
\Lambda_{T}^{\phantom{T} A} = \frac{1}{m}p_\mu (\tilde{e}^A)^\mu = \frac{p^A}{m}\,, \label{constraint:Lambda}
\end{equation}
which shows explicitly that it constrains the three velocity parameters, say $\phi^{4,5,6}$, of the Lorentz transformation 
that relates the tetrad carried by the particle to the background tetrad. Since $\Lambda_T^{\phantom{T} T}$ is fully determined 
by $\Lambda_T^{\phantom{T} I}$, only three of the four constraints given in Eqs.~\eqref{constraint:chi} or~\eqref{constraint:Lambda} are independent\footnote{One can also
see this from the fact that $\boldsymbol{\chi}$ is orthogonal to the timelike vector $\boldsymbol{e}_{T}+\boldsymbol{p}/(m c)$. 
Hence only its three spacelike components are independent.}. We will take the spatial components $\chi_i = 0$ as our three independent constraints on the coordinates $\phi^a$.  

In summary, for the generalized NW SSC, the vector of constraints is
\begin{equation}
\boldsymbol{\xi} \equiv(V^1,V^2,V^3,\chi_1,\chi_2,\chi_3)\,.
\end{equation}
In principle, the computation of the matrix $\boldsymbol{C}$ defined in Eq.~\eqref{CIJ} 
can be performed directly using the
unconstrained symplectic algebra \eqref{xp}, \eqref{xx}, \eqref{SS},
\eqref{Sp}, \eqref{Sx}, \eqref{lambdax}, \eqref{lambdaP}, \eqref{lambdaS}. However, since the constraints are formulated in terms
of the momentum four-vector $p_\mu$ rather than the conjugate momenta $P_i$ and the Hamiltonian $H$, 
it turns out to be quite useful to first compute Poisson brackets
between $p_\mu$ and other phase space quantities, and then make use 
of these results to  compute the matrix $\boldsymbol{C}$. The relevant Poisson brackets
are
\begin{subequations}\label{otherPBs}
\begin{eqnarray}
\{x^i,p_j\} &=& \delta^i_j \,,\\
\{x^i,p_t\} &=& -{v^i} \,,\\
\{x^i,m\} &=& -\frac{1}{m}(p^i - p^tv^i) \,,\\
\{p_i,p_j\} &=& -\frac12 R_{ij\mu\nu}S^{\mu\nu} \,,\\
\{p_i,p_t\} &=& \frac12 R_{ik\mu\nu}v^k S^{\mu\nu} - \Gamma^\mu_{i\nu}p_\mu u^\nu \,,\\
\{p_i,m\} &=& -\frac{1}{m} p_\mu \Gamma^\mu_{ik}(p^k - p^tv^k)\,, \\
\{p_t,m\} &=& \frac{1}{m} p_\mu \Gamma^\mu_{k \nu} (p^\nu v^k-p^ku^\nu)\,,\\
\{p_i,(e_T)_j\} &=& -\Gamma^\mu_{ij} (e_T)_\mu \,, \\
\{p_t,(e_T)_j\} &=& \frac{1}{m} \left (-\Omega_{j \nu} p^\nu + \Gamma^{\mu}_{jk} p_\mu v^k \right )\,,\\
\{S^{\mu\nu},p_i\} &=& 2S^{\lambda[\mu}\Gamma^{\nu]}_{i\lambda} \,, \\
\{S^{\mu\nu},p_t\} &=& -2p^{[\mu}u^{\nu]} - 2S^{\lambda[\mu}\Gamma^{\nu]}_{\lambda k}{v^k} \,, \\
\{S^{\mu\nu},m\} &=&  2p^{[\mu}u^{\nu]}\,, \\
\{S^{\mu\nu},(e_T)_j\} &=& 2 \delta^{[\mu}_j e^{\nu]}_T \,, \\
\{(e_T)_i,m\} &=& - \frac{1}{m^2} p^t \left [p^\nu \Omega_{i\nu} + \right. \nonumber \\
&& \left . p_\mu \Gamma^{\mu}_{ij} (p^j - p^tv^j)\right ]\,, \\
\{(\tilde{e}_T)^\mu,m\} &=& - \frac{1}{m}(\tilde{e}_T)^\mu_{,k}(p^k - p^t v^k)\,,
\end{eqnarray}
\end{subequations}
where the Poisson bracket between an arbitrary phase space function $A$ and the quantity $p_t$ is obtained as follows
\begin{eqnarray}
\{A,p_t\} &=& \{A,-H - \frac12\eta^{AB}(\tilde{e}_A)_\alpha(\tilde{e}_B)_{\beta;t}S^{\alpha\beta}\}\,, \nonumber \\
&=& \frac{\partial A}{\partial t} - \frac{dA}{dt} - \frac12 \eta^{AB}\{A,(\tilde{e}_A)_\alpha(\tilde{e}_B)_{\beta;t}S^{\alpha\beta}\}\,. \nonumber \\
\end{eqnarray}
The total time derivative ${dA}/{dt}$ is then evaluated with the help of the unconstrained equations of motion.
The Poisson brackets~\eqref{otherPBs} along with Eqs.~\eqref{pu_relation} and~\eqref{SS} yield 
\begin{eqnarray}
\{V^i,V^j\}&=& \omega_\mu\, \omega^\mu\, S^{ij} + {\cal O}(S^2)\,,\\
\{V^i,\chi_j\}&=& \frac{\omega_\mu p^\mu}{m}\left(\delta^i_j - \frac{p^i\omega_j}{\omega_\mu p^\mu}\right) + {S^{i\lambda} \tilde{e}^T_{\lambda;\nu}}\left(\delta^\nu_j + \frac{p^\nu p_j}{m^2}\right) \nonumber \\
&& + {\cal O}(S^2)\,,\\
\{\chi_i,\chi_j\}&=&\frac{1}{2m^4}\Big(p_i R_{j\lambda \mu\nu} - p_j R_{i\lambda\mu\nu}\Big)p^\lambda S^{\mu\nu} \nonumber \\
&& -\frac{1}{2m^2}\,R_{ij\mu\nu}\,S^{\mu\nu}  +{\cal O}(S^2)\,,
\end{eqnarray}
The remainders scaling as the square of the particle's spin are dropped, since the pole-dipole particle model is valid
only at linear order in the particle's spin. The matrix $\boldsymbol{C}$ defined in Eq.~\eqref{CIJ} is therefore given by
\begin{equation} \label{Cmatrix}
\boldsymbol{C}=\boldsymbol{K}+\boldsymbol{\Sigma}+{\cal O}(S^2)
\end{equation}
where the matrices $\boldsymbol{K}$ and $\boldsymbol{\Sigma}$ are defined as
\begin{equation}
\boldsymbol{K}=\left( \begin{array}{cc}
\mathbb{O}_3& \boldsymbol{Q}  \\
-\boldsymbol{Q}^T & \mathbb{O}_3  \\
\end{array} \right)\,,
\end{equation}
with
\begin{equation}
Q^{i}_{\phantom{j}j}=\frac{\omega_\mu p^\mu}{m}\left(\delta^i_j - \frac{p^i\omega_j}{\omega_\mu p^\mu}\right)\,,
\end{equation}
and
\begin{widetext}
\begin{equation}\label{matrix:L}
\Sigma^{ij}\equiv\left( \begin{array}{cc}
\omega_\mu \omega^\mu S^{ij} & {S^{i\mu}} \tilde{e}^{T}_{\mu;\nu}\left(\delta^\nu_j + \frac{p^\nu p_j}{m^2}\right)\\
- {S^{j\mu}}\tilde{e}^{T}_{\mu;\nu}\left(\delta^\nu_i + \frac{p^\nu p_i}{m^2}\right) &-\frac{1}{2m^2}R_{k\lambda\mu\nu}S^{\mu\nu}\left[\delta^k_i \delta^\lambda_j + \frac{p^\lambda}{m^2}(\delta^k_i p_j - \delta^k_j p_i)\right] \\
\end{array} \right)\,.
\end{equation}
\end{widetext}
The inverse matrix $\boldsymbol{C}^{-1}$ can be easily  computed at linear order in the spin, the result being
\begin{equation} \label{inverseC}
\boldsymbol{C}^{-1}= \boldsymbol{K}^{-1}-\boldsymbol{K}^{-1} \boldsymbol{\Sigma} \boldsymbol{K}^{-1}+{\cal O}(S^2)\,,
\end{equation}
where 
\begin{equation}\label{matrix:Kinv}
\boldsymbol{K}^{-1}=\left( \begin{array}{cc}
\mathbb{O}_3& -(\boldsymbol{Q}^{-1})^T  \\
\boldsymbol{Q}^{-1} & \mathbb{O}_3  \\
\end{array} \right)\,,
\end{equation}
with
\begin{equation}\label{matrix:Qinv}
[Q^{-1}]^{i}_{\phantom{j}j}= \frac{m}{\omega_\mu p^\mu}\left(\delta^{i}_{j}+\frac{\omega_j p^i}{\omega_t p^t}\right)\,.
\end{equation}
To compute the Dirac brackets between two phase space functions, one also needs the Poisson brackets
between those phase space functions and the constraints. For our purposes, the relevant brackets are given by
\begin{subequations}\label{xPS_constraint_brackets}
\begin{eqnarray}\label{xv}
\{x^i,V^j\}&=&-S^{ij}-S^{jt}\frac{p^i}{p^t} +{\cal O}(S^2)\,,\\
\{x^i,\chi_j\}&=&-\frac{1}{m}\left[\delta^i_j + \frac{p_j}{m^2}(p^i - p^tu^i)\right]+{\cal O}(S^2)\,,\nonumber \\ \,\label{xchi} 
\end{eqnarray}
\begin{eqnarray}
\{P_i,V^j\}&=& p_A S^{j\alpha}\tilde{e}^{A}_{\alpha,i}-S^{jt}\Gamma^\mu_{\nu i} \frac{p^\nu p_\mu}{p^t} +{\cal O}(S^2)\nonumber \\
&=& p_A \tilde{e}^{A}_{k,i}\left(S^{jk}-S^{jt}\frac{p^k}{p^t}\right) +{\cal O}(S^2)\,, \label{Pv} \\ 
\{P_i,\chi_j\}&=& -\frac{1}{m}p_A\tilde{e}^A_{j,i} +\frac{2p_j}{m^3}p^{[\mu} u^{\nu]} \times \nonumber \\
&& \left(E_{i\mu\nu} - p_\lambda\Gamma^\lambda_{i\mu}\delta^t_\nu \right) + S^{\mu\nu}\Big[\frac{1}{2m^2}R_{ij\mu\nu} \nonumber \\
&& - \frac{1}{m}(E_{i\mu\nu;j} + \Gamma^\lambda_{ij} E_{\lambda ij})\Big]   +{\cal O}(S^2)\,, \nonumber \\
\end{eqnarray}
\begin{eqnarray}
\{S^{AB},V^i\}&=& S^{A i} \omega^B - S^{B i}\omega^A +{\cal O}(S^2)\,,\label{SV} \\
\{S^{AB},\chi_i\}&=&  -\frac{2}{m} p^{[A} \tilde{e}^{B]}_i + \frac{2p_i}{m^3}\,p^{[A}u^{B]} \nonumber \\
&& -\frac{2}{m}S^{\mu [A}\Gamma^{B]}_{i\mu}  + {\cal O}(S^2)\label{Schi}\,.
\end{eqnarray}
\end{subequations}
The matrix~\eqref{inverseC} and Eqs.~\eqref{xPS_constraint_brackets}, together
with the unconstrained algebra given by Eqs.~\eqref{xPalgebra_unconstrained} and~\eqref{Salgebra_unconstrained}, is all one 
needs to compute the Dirac brackets according to Eq.~\eqref{DB}. Our results for the Dirac brackets involving $x^i$ and $P_j$ are given by
\begin{widetext}
\begin{subequations}\label{xPDBs}
\begin{eqnarray}
\{x^i,x^j\}_{\rm DB} &=& \left[\frac{\omega^\mu \omega_\mu-2 p^\nu \omega_\nu}{(p^\sigma \omega_\sigma)^2}\right]\left(S^{ij}-S^{it}\frac{p^j}{p^t}+S^{jt}\frac{p^i}{p^t}\right)+{\cal O}(S^2)={\cal O}(S^2)\,, \\
\{x^i,P_j\}_{\rm DB} &=& \delta^i_j+ \left(S^{ik}-S^{it}\frac{p^k}{p^t}+S^{kt}\frac{p^i}{p^t}\right)
\left[\frac{\omega^\mu \omega_\mu-2 p^\nu \omega_\nu}{(p^\sigma \omega_\sigma)^2}\right]p_{\alpha}\tilde{e}^\alpha_A
\tilde{e}^A_{k,j}+{\cal O}(S^2)=\delta^i_j+ {\cal O}(S^2)\,,\\
\{P_i,P_j\}_{\rm DB} &=& \left(S^{kl}-S^{kt}\frac{p^l}{p^t}+S^{lt}\frac{p^k}{p^t}\right)\left[\frac{\omega^\mu \omega_\mu-2 p^\nu \omega_\nu}{(p^\sigma \omega_\sigma)^2}\right] p_{\alpha}\tilde{e}^\alpha_A
\tilde{e}^A_{k,i}\, p_{\beta}\tilde{e}^\beta_B
\tilde{e}^B_{l,j}+{\cal O}(S^2)={\cal O}(S^2)\,.
\end{eqnarray}
\end{subequations}
\end{widetext}
The crucial point now is that Eq.~\eqref{eq:omegaNW} implies $\omega^\mu \omega_\mu=2 p^\mu \omega_\mu$, and therefore all terms linear in the particle's spin on the right-hand side of Eqs.~\eqref{xPDBs} vanish. Hence the Dirac bracket algebra between $x^i$ and $P_j$ is canonical up to terms quadratic in the particle's spin.

The Dirac brackets involving the spin variables are most effectively  computed by considering the projection of the spin tensor onto the spacelike 
background tetrad vectors, \textit{i.e.} $S^{IJ} = S^{\mu\nu}\tilde{e}^I_\mu \tilde{e}^J_\nu$. We find 
\begin{widetext}
\begin{subequations}\label{SDBs}
\begin{eqnarray}\label{DB:XS}
\{x^i,S^{KL}\}_{\rm DB} &=& \left[\frac{\omega^\mu \omega_\mu-2 p^\nu \omega_\nu}{(p^\sigma \omega_\sigma)^2}\right]\left(S^{i\alpha}+S^{\alpha t}\frac{p^i}{p^t}\right)p^\beta\left(\tilde{e}_\beta^K \tilde{e}_\alpha^L - \tilde{e}_\beta^L \tilde{e}_\alpha^K \right)+{\cal O}(S^2)={\cal O}(S^2)\\
\{P_i,S^{KL}\}_{\rm DB}&=&\left[\frac{\omega^\mu \omega_\mu-2 p^\nu \omega_\nu}{(p^\sigma \omega_\sigma)^2}\right]
\left(S^{\gamma k}-S^{\gamma t}\frac{p^k}{p^t}\right)p^\alpha p_\beta\tilde{e}^\beta_C \tilde{e}^C_{k,i}
\left(\tilde{e}_\alpha^L \tilde{e}_\gamma^K - \tilde{e}_\gamma^L \tilde{e}_\alpha^K\right)+{\cal O}(S^2)={\cal O}(S^2)\,, \\
\{S^{IJ},S^{KL}\}_{\rm DB} &=& \left[\frac{\omega^\mu \omega_\mu-2 p^\nu \omega_\nu}{(p^\sigma \omega_\sigma)^2}\right]S^{\gamma\delta}
p^\alpha p^\beta\left(\tilde{e}_\alpha^L \tilde{e}_\delta^K - \tilde{e}_\delta^L \tilde{e}_\alpha^K\right)
\left(\tilde{e}_\beta^J \tilde{e}_\gamma^I - \tilde{e}_\gamma^J \tilde{e}_\beta^I\right) + S^{IK}\delta^{JL}+S^{JL}\delta^{IK}\nonumber \\
 &&-S^{IL}\delta^{JK}-S^{JK}\delta^{IL} +{\cal O}(S^2)\nonumber \\ &=&S^{IK}\delta^{JL}+S^{JL}\delta^{IK}
-S^{IL}\delta^{JK}-S^{JK}\delta^{IL} +{\cal O}(S^2)\,,
\label{DB:SS}
\end{eqnarray}
\end{subequations}
\end{widetext}
where we have used $\omega^I = p^I$, which follows directly from Eq.~\eqref{eq:omegaNW}. Again the terms proportional to $\omega_\mu\omega^\mu - 2p^\nu\omega_\nu$ disappear. Defining a three-dimensional spin vector by
\begin{equation}\label{canonicalspin}
S^I=\frac12 \epsilon^{IJK}\, S^{JK}
\end{equation}
one can immediately rewrite Eqs.~\eqref{SDBs} as
\begin{subequations}\label{SDBalgebra}
\begin{eqnarray}
\{x^i,S^J\}_{\rm DB} &=& {\cal O}(S^2)\,,\\
\{P_i,S^J\}_{\rm DB} &=& {\cal O}(S^2)\,,\\
\{S^I,S^J\}_{\rm DB} &=& \epsilon_{IJK} S^K + {\cal O}(S^2)\label{eq:SiSj}\,.
\end{eqnarray}
\end{subequations}
Equations~\eqref{SDBalgebra} imply that the phase-space variables $\{x^i,P_j,S^K\}$ provided by the generalized NW 
SSC are canonical at linear order in the particle's spin.
 
\subsection{Hamiltonian in the generalized Newton-Wigner SSC}
\label{sec:hamiltonian}

In this section, we provide an explicit expression for the Hamiltonian~\eqref{Hamiltonian} in
the NW SSC, at linear order in the particle's spin. As explained in Sec.~\ref{sec:DBprimer}, this is simply
obtained by inserting the NW SSC directly into the unconstrained
Hamiltonian. Also, we express this constrained Hamiltonian in
terms of the variables $x^i$, $P_j$, $S^K$, which have been proven 
in Sec.~\ref{sec:NWDB} to be canonical at linear order
in the particle spin.

We begin by rewriting the quantity $p_t$ appearing in the unconstrained Hamiltonian~\eqref{Hamiltonian} in terms of the mass $m=\sqrt{-p_\mu p^\mu}$ and the spatial components $p_i$ of the momentum four-vector. The result is
\begin{equation}\label{pt_solved}
p_t=-\beta^i p_i-\alpha \sqrt{m^2 + \gamma^{ij}p_i p_j}\,,
\end{equation}
where 
\begin{subequations}
\begin{eqnarray}
\alpha &=& \frac{1}{\sqrt{-g^{tt}}}\,,\\
\beta^i &=& \frac{g^{ti}}{g^{tt}}\,,\\
\gamma^{ij} &=& g^{ij}-\frac{g^{ti} g^{tj}}{g^{tt}}\,.
\end{eqnarray}
\end{subequations}
The crucial usefulness of Eq.~\eqref{pt_solved} resides in the fact that the canonical phase-space variables $\{x^i,P_j,S^K\}$ have vanishing Dirac brackets with the mass at linear order in the particle spin. We have established this result by explicit computation. As an illustration, we provide the details of the computation of the Dirac bracket between $x^i$ and the mass (the other brackets involving the mass are  computed in a similar fashion). We start from
\begin{eqnarray}
\{x^i,m\}_{\rm DB} &=& \{x^i,\sqrt{-p_\mu p^\mu}\}_{\rm DB}\,, \nonumber \\
&=& -\frac{1}{2m}\{x^i,g^{\mu\nu}p_\mu p_\nu\}_{\rm DB}\,, \nonumber \\
&=& -\frac{1}{m}p^\mu\{x^i,p_\mu\}_{\rm DB}\,,
\end{eqnarray}
the last line following from $\{x^i,x^j\}_{\rm DB} = {\cal O}(S^2)$. Using Eq.~\eqref{capP4v} together with the fact that the Dirac bracket with the Hamiltonian gives the constrained equations of motion yields
\begin{eqnarray}
\{x^i,m\}_{\rm DB} &=& -\frac{1}{m}p^\mu\{x^i,P_\mu - E_{\mu\alpha\beta}S^{\alpha\beta}\}_{\rm DB}\,, \nonumber \\
&=& -\frac{1}{m}(p^i - p^t v^i) + \frac{1}{m}p^\mu E_{\mu\alpha\beta}\{x^i,S^{\alpha\beta}\}_{\rm DB} \,, \nonumber \\ \, \label{ximDB}
\end{eqnarray} 
where Eq.~\eqref{pu_relation} must be employed to express the four-velocity components $v^i$ in terms of canonical variables. Substituting Eq.~\eqref{eq:omegaNW} into Eq.~\eqref{pu_relation}, it is straightforward to show that
\begin{equation}
p^i - p^t v^i = -\frac{m}{\omega_\nu p^\nu}\tilde{e}^T_{\lambda;\sigma}p^\sigma\left(S^{i\lambda} - S^{t\lambda}\frac{p^i}{p^t}\right)\,. \label{pi-ptvi}
\end{equation}
Next the Dirac bracket between $x^i$ and $S^{\alpha\beta}$ at linear order in spin can be  computed directly following the procedure outlined in Sec.~\ref{sec:NWDB}, the result being
\begin{equation}
\{x^i,S^{\alpha\beta}\}_{\rm DB} = -\frac{2m(\tilde{e}^T)^{[\alpha}}{\omega_\nu p^\nu}
\left(S^{\beta]i} + S^{\beta]k}\frac{\omega_k p^i}{\omega_t p^t}\right)\,. \label{xiSalphabetaDB}
\end{equation}
Hence, since $E_{\mu\alpha\beta}$ is antisymmetric in $\alpha \leftrightarrow \beta$, we get
\begin{eqnarray}
 E_{\mu\alpha\beta}\{x^i,S^{\alpha\beta}\}_{\rm DB} &=& E_{\mu\alpha\beta}\frac{2m(\tilde{e}^T)^{\alpha}}{\omega_\nu p^\nu}\left(S^{\beta i} + S^{\beta k}\frac{\omega_k p^i}{\omega_t p^t}\right) \nonumber \\
&=& \frac{m}{\omega_\nu p^\nu}(\tilde{e}^T)_{\beta;\mu}\left(S^{\beta i} + S^{\beta k}\frac{\omega_k p^i}{\omega_t p^t}\right)\,,\nonumber \\ \, \label{Econtract}
\end{eqnarray}
the second line following  from the definition $2E_{\mu\alpha\beta} = \eta^{AB}(\tilde{e}_A)_\alpha(\tilde{e}_B)_{\beta;\mu}$. Substituting Eqs.~\eqref{pi-ptvi},~\eqref{xiSalphabetaDB} and~\eqref{Econtract} into Eq.~\eqref{ximDB} one obtains
\begin{eqnarray}
\{x^i,m\}_{\rm DB} &=& \frac{1}{\omega_\nu p^\nu}\bigg[(\tilde{e}^T)_{\lambda;\sigma}p^\sigma \left(S^{i\lambda} - S^{t\lambda}\frac{p^i}{p^t}\right) \nonumber \\
&& + p^\mu (\tilde{e}^T)_{\beta;\mu} \left(S^{\beta i} + S^{\beta k}\frac{\omega_k p^i}{\omega_t p^t}\right)\bigg] \,.
\end{eqnarray}
Renaming dummy indices and making use of the NW SSC to rewrite $S^{t\lambda} = - S^{k\lambda}\omega_k/\omega_t$, one can see that all terms cancel, therefore showing that the mass commutes with $x^i$ under the Dirac brackets.  

Since the constrained Hamiltonian depends only on $\{x^i,P_j,S^K\}$ and the mass $m$, it follows that the mass may be treated as a constant when taking the Dirac bracket between an arbitrary function of constrained phase-space variables and the Hamiltonian.

Our Hamiltonian~\eqref{Hamiltonian} now takes the form
\begin{eqnarray}\label{Hamiltonian_canonical}
\bar{H} &=& \beta^i p_i + \alpha\, \sqrt{m^2 + \gamma^{ij}p_i p_j}
- E_{tAB}S^{AB}\,.
\end{eqnarray}
Equation~\eqref{canonicalspin} implies $S^{IJ}=\epsilon^{IJK} S^K$, while the 
NW SSC [Eqs.~\eqref{eq:NW} and~\eqref{eq:omegaNW}] implies
\begin{equation}\label{NWSSC_projected}
S^{TI}=\frac{S^{IJ}\omega_{J}}{\omega_{T}}\,,
\end{equation}
where
\begin{subequations}
\begin{eqnarray}
\omega_{T} &=& \omega_\mu\, \tilde{e}^\mu_{T}=p_\mu\, \tilde{e}^\mu_{T}-m \,,\\
\omega_{I} &=& \omega_\mu\, \tilde{e}^\mu_{I}=p_\mu\, \tilde{e}^\mu_{I}\,.
\end{eqnarray}
\end{subequations}
The canonical momenta $P_i$ are related to the linear momenta $p_i$ by Eq.~\eqref{Pic}, which may be rewritten in terms of the canonical spin variables as
\begin{eqnarray}
P_i &=& p_i+ E_{iAB} S^{AB}\,, \nonumber \\
&=& p_i + \left(2E_{iTJ}\frac{\omega_K}{\omega_T} + E_{iJK}\right)\epsilon^{JKL}S^L\,.\label{Pi_canonical}
\end{eqnarray}
In principle, in order to express the Hamiltonian~\eqref{Hamiltonian_canonical} in terms of the
canonical momenta $P_i$, one must invert Eq.~\eqref{Pi_canonical} to obtain $p_i$ as function of canonical variables (recall that $\omega_\mu$ depends on $p_\mu$). However, because our Hamiltonian is valid only at linear order in the test-particle's spin, it is sufficient to write
\begin{equation}\label{Pi_canonical2}
p_i=P_i- \left(2E_{iTJ}\frac{\bar{\omega}_K}{\bar{\omega}_T} + E_{iJK}\right)\epsilon^{JKL}S^L+{\cal O}(S^2)\,,
\end{equation}
where
\begin{subequations}
\begin{eqnarray}
\bar{\omega}_\mu&=&\bar{P}_\mu-m \,\tilde{e}^{T}_\mu\\
\bar{P}_i &=& P_i\,,\\
\bar{P}_t &=& -\beta^i\,P_i-\alpha\, \sqrt{m^2 +\gamma^{ij}\,P_i\, P_j}\,,\\
\label{bomegaT}
\bar{\omega}_T &=& \bar{\omega}_\mu\,\tilde{e}^\mu_{T}=\bar{P}_\mu\,\tilde{e}^\mu_{T}-m\,,\\
\bar{\omega}_I &=&\bar{\omega}_\mu\,\tilde{e}^\mu_{I}= \bar{P}_\mu\,\tilde{e}^\mu_{I}\,.
\label{bomegaI}
\end{eqnarray}
\end{subequations}
We may now write  the constrained Hamiltonian~\eqref{Hamiltonian_canonical} as
\begin{eqnarray}
\bar{H} &=& \beta^i\,p_i+\alpha\,\sqrt{m^2 +\gamma^{ij}\,p_i\, p_j} - F_t^KS_K +{\cal O}(S^2)\,, \nonumber \\ \label{Hamiltonian_canonical_final}
\end{eqnarray}
where
\begin{equation}\label{FmuK_def}
F_\mu^K = \left(2E_{\mu TI}\frac{\bar{\omega}_J}{\bar{\omega}_T} + E_{\mu IJ}\right)\epsilon^{IJK}\,.
\end{equation}
By substituting expression~\eqref{Pi_canonical2} for $p_i$ into Eq.~\eqref{Hamiltonian_canonical_final} and expanding to linear order in spin, one arrives at last at the following Hamiltonian
\begin{eqnarray}\label{Hamiltonian_canonical_final2}
\bar{H} &=& \bar{H}_{\rm NS} - \left(\beta^iF_i^K + F_t^K + \frac{\alpha \gamma^{ij}P_i F_j^K}{\sqrt{m^2 + \gamma^{ij}P_iP_j}}\right)S_K \,,\nonumber \\ \label{Hexpanded}
\end{eqnarray}
where $\bar{H}_{\rm NS}$ is the Hamiltonian for a non-spinning particle, simply given by
\begin{equation}\label{eq:Hns}
\bar{H}_{\rm NS} = \beta^iP_i + \alpha \sqrt{m^2 + \gamma^{ij}P_iP_j}\,.
\end{equation}

\section{Explicit Hamiltonian for specific background spacetimes}\label{sec:explicit}

\subsection{Spherically symmetric spacetime in isotropic coordinates }
\label{sec:SS_ISO_Hamiltonian}

The line element for a generic spherically symmetric spacetime in isotropic coordinates is given by
\begin{equation}
ds^2 = -f(\rho) \,dt^2 + h(\rho)(dx^2+dy^2+dz^2)\,,
\end{equation}
where $\rho^2 = x^2 + y^2 + z^2$. The natural tetrad associated with this spacetime and coordinate system is
\begin{subequations}\label{SStetrad_iso}
\begin{eqnarray}
\tilde{e}_T^\mu &=& \frac{1}{\sqrt{f}}\delta^\mu_0\,, \\
\tilde{e}_I^\mu &=& \frac{1}{\sqrt{h}} \delta^\mu_I \,,
\end{eqnarray}
\end{subequations}
where the symbol $\delta^\mu_I$ is equal to 0 when $\mu = 0$ and it is equal to 1 when $\mu = I$ numerically\footnote{More precisely, even though the spacetime index $\mu$ and the internal tetrad index $I$ are completely different in character, both indices may take on the same numerical value (1,2 or 3 associated with $x,y$ and $z$ respectively).}. With a metric and a convenient tetrad in hand, one may now compute the quantity $E_{\mu AB}$ as follows

\begin{equation}
E_{\mu AB} = -\frac{1}{2}\left[(\tilde{e}_A)_\lambda (\tilde{e}^\lambda_B)_{,\mu} + (\tilde{e}_A)_\lambda \Gamma^\lambda_{\mu \gamma} \tilde{e}^\gamma_B\right]\,.
\end{equation}  
The algebra is straightforward and the result is
\begin{subequations}\label{EmuAB_Schw_ISO}
\begin{eqnarray}
E_{\mu TI} &=&  \frac{f^\prime}{4\sqrt{fh}} \delta_{\mu}^0 \, n_I \,,\\
E_{\mu JK} &=&  -\frac{h^\prime}{2h} \, \delta_{\mu[J} n_{K]} \,,
\end{eqnarray}
\end{subequations}
where the prime symbol denotes a derivative with respect to $\rho$, and where $n_I = (x/\rho,y/\rho,z/\rho)$. The last ingredients needed in order to obtain the explicit Hamiltonian are $\bar{\omega}_T$ and $\bar{\omega}_K$ defined in 
Eqs.~(\ref{bomegaT}), (\ref{bomegaI}). A quick computation yields
\begin{subequations}\label{tildeomega_Schw_ISO}
\begin{eqnarray}
\bar{\omega}_T &=&  - \sqrt{m^2 + \gamma^{ij}P_iP_j} - m\,, \nonumber \\
&\equiv& -m\left(1 + \sqrt{Q}\right) \,, \\
\bar{\omega}_K &=& \frac{1}{\sqrt{h}}\, P_K \,,
\end{eqnarray}
\end{subequations}
where $Q = 1 + \gamma^{ij} P_iP_j/m^2$ and $P_K = P_j \delta^j_K$. By substituting Eqs.~\eqref{EmuAB_Schw_ISO} and~\eqref{tildeomega_Schw_ISO} into Eq.~\eqref{FmuK_def}, we obtain the following expression for the quantity $F_\mu^I$ 
\begin{subequations}\label{FmuK_SS_ISO}
\begin{eqnarray}
F_0^I &=& -\frac{1}{m(1+\sqrt{Q})}\frac{f^\prime}{2\sqrt{f}h}\epsilon^{IJK}n_JP_K  \,,\\
F_j^I &=&  -\frac{h^\prime}{2h} \epsilon^{IJK}\delta_{Jj}n_K\,.
\end{eqnarray}
\end{subequations}
Finally, by substituting Eq.~\eqref{FmuK_SS_ISO} into the Hamiltonian~\eqref{Hexpanded} and performing simple algebra, we arrive at

\begin{equation}\label{HSS_iso}
\bar{H} = \bar{H}_{\rm NS} + \left[\frac{\sqrt{Q}(f^\prime h - fh^\prime) - fh^\prime}{2M\rho \sqrt{f}h^2\sqrt{Q}(1+\sqrt{Q})}\right](\boldsymbol{L}\cdot\boldsymbol{S}^\ast),
\end{equation}
where $\bar{H}_{\rm NS}$ is the Hamiltonian for a non-spinning
particle, and where
\begin{eqnarray}
Q &=& 1 + \frac{1}{h}\hat{\boldsymbol{P}}^2 \,,\\
\hat{\boldsymbol{P}}^2 &=& \delta^{JK}\frac{P_JP_K}{m^2} = \delta^{jk}\frac{P_jP_k}{m^2}\,, \label{hatPsquare_ISO}\\
\boldsymbol{L}\cdot\boldsymbol{S}^\ast &=& \rho\, \epsilon^{IJK}\,n_I\,P_J\,\left(\frac{M \,S_K}{m}\right) \,. \label{LdotS_ISO}
\end{eqnarray}
The quantity $M$ in Eqs.~(\ref{HSS_iso}) and (\ref{LdotS_ISO}) is introduced in anticipation of specialization
to the Schwarzschild metric below.  Since a spherically symmetric
spacetime possesses an $SO(3)$ symmetry (associated with rotation of
the $x,y,z$ coordinates among themselves) that is shared by the
internal tetrad space, one may accompany any coordinate rotation by
the corresponding tetrad rotation, thereby preserving the functional
form of the Hamiltonian~\eqref{HSS_iso}, as well as the
quantities~\eqref{hatPsquare_ISO} and~\eqref{LdotS_ISO}. Thus one may
meaningfully identify the vectors $L_I = \rho \, \epsilon^{IJK}n_JP_K$
and $S_I$ (which really live in the tetrad internal space) with
spacetime vectors $L_i$ and $S_i$ which transform accordingly under
rotations of the coordinates $x,y,z$.  

In the limit of flat spacetime, the Hamiltonian reduces to $\bar{H}_{\rm NS}$ as expected, since the Cartesian components of the spin are all constants of motion. For the Schwarzschild spacetime in isotropic coordinates, we have
\begin{equation}
ds^2 = -\left[\frac{1 - M/(2\rho)}{1 + M/(2\rho)}\right]^2dt^2 + \left(1+\frac{M}{2\rho}\right)^4(dx^2+dy^2+dz^2)\,.
\end{equation}
Substituting these explicit expressions for $f(\rho)$ and $h(\rho)$ in the Hamiltonian~\eqref{HSS_iso}, one finds
\begin{multline}
\bar{H} = \bar{H}_{\rm NS} + \frac{\psi^6}{\rho^3 \sqrt{Q}(1+\sqrt{Q})}\\\times\left[1-\frac{M}{2\rho} + 2\left(1-\frac{M}{4\rho}\right)\sqrt{Q}\right](\boldsymbol{L}\cdot\boldsymbol{S}^\ast)\,,
\end{multline}
where $\psi = (1 + M/2\rho)^{-1}$\,.

\subsection{Spherically symmetric spacetime in spherical coordinates}

In this case, the metric takes the form
\begin{equation}\label{SSmetric_sph}
ds^2 = -f(r)dt^2 + h(r)dr^2 + r^2 d\theta^2 + r^2\sin^2\theta d\phi^2\,.
\end{equation}
Note that the functions $f$ and $h$ appearing above are not the same as in the isotropic case. However we follow here generally accepted notation conventions. The natural tetrad associated with this spacetime and coordinate system is
\begin{subequations}\label{tetradSS_sph}
\begin{eqnarray}
\tilde{e}_T^\mu &=& \frac{1}{\sqrt{f}}\delta^\mu_{t}\,, \\
\tilde{e}_1^\mu &=& \frac{1}{\sqrt{h}} \delta^\mu_{r} \,, \\
\tilde{e}_2^\mu &=& \frac{1}{r} \delta^\mu_{\theta} \,, \\
\tilde{e}_3^\mu &=& \frac{1}{r\sin\theta} \delta^\mu_{\phi} \,. 
\end{eqnarray}
\end{subequations}
The metric~\eqref{SSmetric_sph} and the tetrad~\eqref{tetradSS_sph} then lead to the following result
\begin{subequations}\label{EmuAB_sph}
\begin{eqnarray}
E_{tAB} &=& \frac{f^\prime}{2\sqrt{fh}}\delta_{[A}^{T}\delta_{B]}^1 \,, \\
E_{rAB} &=& 0 \,, \\
E_{\theta AB} &=& \frac{1}{\sqrt{h}}\delta_{[A}^{1}\delta_{B]}^2 \,, \\
E_{\phi AB} &=& \frac{\sin\theta}{\sqrt{h}}\delta_{[A}^{1}\delta_{B]}^3 + \cos\theta\,\delta_{[A}^{2}\delta_{B]}^3\,.
\end{eqnarray}
\end{subequations}
Next the computation of $\bar{\omega}_T$ and $\bar{\omega}_K$ yields
\begin{subequations}\label{tilde_omega_sph}
\begin{eqnarray}
\bar{\omega}_T &=& -m\left(1+\sqrt{Q}\right)\,, \\
\bar{\omega}_1 &=& \frac{1}{\sqrt{h}}P_r \,, \\
\bar{\omega}_2 &=& \frac{1}{r}P_\theta \,, \\
\bar{\omega}_3 &=& \frac{1}{r\sin\theta}P_\phi \,,
\end{eqnarray}
\end{subequations}
Equations~\eqref{EmuAB_sph} and~\eqref{tilde_omega_sph} then allow us to obtain $F_\mu^I$. The result is
\begin{subequations}
\begin{eqnarray}
F_\mu^1 &=& \cos\theta \, \delta_\mu^\phi \,,\\
F_\mu^2 &=& \left(\frac{f^\prime}{2r\sin\theta\sqrt{fh}}\right)\left(\frac{\hat{P}_\phi}{1+\sqrt{Q}}\right)\delta_\mu^t \nonumber \\
&&  - \frac{\sin\theta}{\sqrt{h}} \delta_\mu^\phi \,, \\
F_\mu^3 &=& -\left(\frac{f^\prime}{2r\sqrt{fh}}\right)\left(\frac{\hat{P}_\theta}{1+\sqrt{Q}}\right)\delta_\mu^t + \frac{1}{\sqrt{h}}\delta_\mu^\theta\,,\nonumber \\&&
\end{eqnarray}
\end{subequations}
where again $\hat{P}_i = P_i/m$. The Hamiltonian then follows immediately
\begin{eqnarray}
\bar{H} &=& \bar{H}_{\rm NS} + \frac{f^\prime}{2(1+\sqrt{Q})r\sqrt{fh}}\left(-\frac{1}{\sin\theta}\hat{P}_\phi S_2 + \hat{P}_\theta S_3\right) \nonumber \\
&& - \sqrt{\frac{f}{Q}}\left(\frac{\cos\theta}{r^2\sin^2\theta}\hat{P}_\phi S_1 - \frac{\hat{P}_\phi S_2}{r^2\sqrt{h}\sin\theta} + \frac{\hat{P}_\theta S_3}{r^2\sqrt{h}}\right)\,. \nonumber  \\ \label{HSS_sph}
\end{eqnarray}
The spin terms in the first line of the Hamiltonian~\eqref{HSS_sph} are the spherical coordinate equivalent of the $\boldsymbol{L}\cdot\boldsymbol{S}^\ast$ terms of the isotropic Hamiltonian~\eqref{HSS_iso}. The spin terms in the second line of Eq.~\eqref{HSS_sph} do not vanish in the flat space limit $f=h=1$, and therefore represent coordinate effects related to the fact that the components of the spin in spherical coordinates and its associated tetrad must evolve, even in the absence of spin-orbit coupling. Such spin terms in the Hamiltonian represent therefore a type of gauge terms. 

Notice however that one could in principle eliminate these gauge terms in the Hamiltonian by picking a ``Cartesian'' tetrad, even though the coordinate system chosen is the spherical one. For example one could pick the ``isotropic'' tetrad~\eqref{SStetrad_iso}, taking care of transforming the components of $\boldsymbol{\tilde{e}}_A$ from isotropic to spherical coordinates. In that case the spin degrees of freedom $S_K$, which live in the internal tetrad space, behave as the components of the spin in Cartesian coordinates, and in that case the flat space limit of the Hamiltonian should be free of gauge terms and should reduce to the non-spinning Hamiltonian.

For the Schwarzschild spacetime, $f = 1/h = 1 -{2M}/{r}$, and we obtain
\begin{eqnarray}\label{H:schwarzschild}
\bar{H} &=& \bar{H}_{\rm NS} + \frac{M}{r^3(1+\sqrt{Q})} \left(-\frac{1}{\sin\theta}\hat{P}_\phi S_2 + \hat{P}_\theta S_3\right) \nonumber \\
&& -\frac{1-{2M}/{r}}{\sqrt{Q}}\bigg[\frac{\cos\theta}{r^2\sin^2\theta}\left(1-\frac{2M}{r}\right)^{-1/2} \hat{P}_\phi S_1 \nonumber \\
&& - \frac{\hat{P}_\phi S_2}{r^2\sin\theta} + \frac{\hat{P}_\theta S_3}{r^2}\bigg]\,,
\end{eqnarray}
where
\begin{equation}
Q = 1 + \left(1-\frac{2M}{r}\right)\hat{P}_r^2 + \frac{1}{r^2}\hat{P}_\theta^2 + \frac{1}{r^2\sin^2\theta}\hat{P}_\phi^2\,.
\end{equation}

 \subsection{Kerr spacetime in Boyer-Lindquist coordinates}
\label{sec:Kerr_Hamiltonian}

Not surprisingly the computation of the Hamiltonian is much more involved in Kerr spacetime, whose line element, in Boyer-Lindquist coordinates, is given by
\begin{eqnarray}\label{KerrMetric}
ds^2 &=& \left(-1 + \frac{2Mr}{\Sigma}\right)\,dt^2 - \frac{4aMr\sin^2\theta}{\Sigma}\,dt\,d\phi \nonumber \\
&& + \frac{\Lambda \sin^2\theta}{\Sigma}\,d\phi^2 + \frac{\Sigma}{\Delta}\,dr^2 + \Sigma\,d\theta^2 \,, 
\end{eqnarray}
where
\begin{subequations}\label{KerrMetric_functions}
\begin{eqnarray}
\Sigma &=& r^2 + a^2\cos^2\theta \,, \\
\Delta &=& r^2 + a^2 - 2Mr \,, \\
\varpi^2 &=& r^2 + a^2 \,, \\
\Lambda &=& \varpi^4 - a^2\Delta\sin^2\theta \,.
\end{eqnarray}
\end{subequations}
For sake of shortening some further formulas, we also introduce the quantity
\begin{equation}
\rho^2 = r^2 - a^2 \cos^2\theta\,.
\end{equation}
Our choice for the reference tetrad  is given by the ``spheroidal'' tetrad
 \begin{subequations}
 \begin{eqnarray}
 \tilde{e}^T_\mu &=& \delta^t_\mu \sqrt{\frac{\Delta \Sigma}{\Lambda }}\,,\\
 \tilde{e}^1_\mu &=& \delta^r_\mu \sqrt{\frac{\Sigma}{\Delta }}\,,\\
 \tilde{e}^2_\mu &=& \delta^\theta_\mu \sqrt{\Sigma}\,, \\
 \tilde{e}^3_\mu &=&  - \frac{2 a M r \sin\theta }{\sqrt{\Lambda \Sigma} } \delta^t_\mu +\delta^\phi_\mu \sin\theta \sqrt{\frac{\Lambda} {\Sigma}}\,,
 \end{eqnarray}
 \end{subequations}
which reduces to the ``spherical'' tetrad \eqref{tetradSS_sph} for $a=0$. This tetrad then leads to the following components for the quantities $E_{\mu \, AB}$
\begin{subequations}
\begin{eqnarray}
E_{t\, T1} &=& \frac{M \varpi^2 \rho^2}{2 \sqrt{\Lambda } \Sigma^2} \,, \\
E_{t\, T2} &=&  -\frac{a^2 \sqrt{\Delta} M r \cos \theta \sin \theta}{\sqrt{\Lambda } \Sigma^2}\,, \\
E_{t\, T3} &=& 0 \,, \\
E_{t\, 12} &=& 0 \,, \\
E_{t\, 13} &=&  \frac{a \sqrt{\Delta} M \rho^2 \sin \theta}{2 \sqrt{\Lambda } \Sigma^2}\,, \\
E_{t\, 23} &=& -\frac{a M r \varpi^2 \cos \theta}{\sqrt{\Lambda } \Sigma^2} \,,
\end{eqnarray}
\end{subequations}
\begin{subequations}
\begin{eqnarray}
E_{r\, T1} &=&0  \,, \\
E_{r\, T2} &=& 0 \,, \\
E_{r\, T3} &=& -\frac{a M \left(2r^2\Sigma + \varpi^2\rho^2\right) \sin \theta}{2 \sqrt{\Delta} \Lambda 
   \Sigma}\,,\\
E_{r\, 12} &=& \frac{a^2 \cos \theta \sin \theta}{2 \sqrt{\Delta} \Sigma} \,, \\
E_{r\, 13} &=& 0 \,, \\
E_{r\, 23} &=& 0 \,,
\end{eqnarray}
\end{subequations}
\begin{subequations}
\begin{eqnarray}
E_{\theta\, T1} &=& 0 \,, \\
E_{\theta\, T2} &=& 0 \,, \\
E_{\theta\, T3} &=& \frac{a^3 \sqrt{\Delta} M r \cos \theta \sin ^2\theta}{\Lambda  \Sigma} \,, \\
E_{\theta\, 12} &=& \frac{\sqrt{\Delta} r}{2 \Sigma} \,, \\
E_{\theta\, 13} &=& 0 \,, \\
E_{\theta\, 23} &=& 0 \,,
\end{eqnarray}
\end{subequations}
\begin{subequations}
\begin{eqnarray}
&& E_{\phi\, T1} = -\frac{a M \sin ^2\theta}{2 \sqrt{\Lambda} \Sigma^2} \left(2r^2\Sigma + \varpi^2\rho^2\right) \,, \\
&& E_{\phi\, T2} = \frac{a^3 \sqrt{\Delta} M r \cos \theta \sin ^3\theta}{\sqrt{\Lambda } \Sigma^2} \,, \\
&& E_{\phi\, T3} = 0 \,, \\
&& E_{\phi\, 12} = 0 \,, \\
&& E_{\phi\, 13} = \frac{\sqrt{\Delta} \sin \theta}{2
   \sqrt{\Lambda } \Sigma^2} \left(r\Sigma^2 - a^2 M \rho^2 \sin^2\theta\right)\!, \\
&&E_{\phi\, 23} = \frac{\left(2 M r \varpi^4+\Delta \Sigma^2\right) \cos \theta}{2 \sqrt{\Lambda } \Sigma^2} \,,
\end{eqnarray}
\end{subequations}
while $\bar{\omega}_T$ and $\bar{\omega}_K$ are easily found to be
\begin{subequations}\label{tildeomega_kerr}
\begin{eqnarray}
\bar{\omega}_T &=&   -m\left(1 + \sqrt{Q}\right) \,, \\
\bar{\omega}_1 &=&  P_r \sqrt{\frac{\Delta}{\Sigma}}\,,\\
\bar{\omega}_2 &=&  P_\theta \sqrt{\frac{1}{\Sigma}}\,,\\
\bar{\omega}_3 &=&  P_\phi \frac{\sqrt{\Sigma}}{\sin\theta \sqrt{\Lambda}}\,,
\end{eqnarray}
\end{subequations}
where
\begin{align}\nonumber
Q =& 1 + \gamma^{ij} {\hat{P}_i\hat{P}_j}\\=&1 +\frac{\Delta}{\Sigma} \hat{P}_r^2+\frac{1}{\Sigma} \hat{P}_\theta^2+\frac{\Sigma}{\Lambda \sin^2\theta} \hat{P}_\phi^2\,,
\end{align}
with $\hat{P}_i \equiv P_i/m$.
The coefficients $F^K_\mu$ are finally given by
\begin{subequations}
\begin{eqnarray}
F_t^1 &=&2 a M r \cos\theta\left[\frac{a \sqrt{\Delta}}{\Lambda (1+\sqrt{Q})\Sigma^{3/2}}\hat{P}_\phi - \frac{\varpi^2}{\sqrt{\Lambda } \Sigma^2}\right] \,, \nonumber\\&&\\
F_r^1 &=& -\frac{a M  \left(2r^2\Sigma + \varpi^2 \rho^2\right) \sin\theta}{\sqrt{\Delta} \Lambda  \left(1+\sqrt{Q}\right) \Sigma^{3/2}}\hat{P}_\theta \,, \\
F_\theta^1 &=& \frac{2 a^3  M r   \cos \theta \sin ^2\theta}{\Lambda(1+  \sqrt{Q} )} \sqrt{\frac{\Delta}{\Sigma^3}}\hat{P}_\theta \,, \\
F_\phi^1 &=&\cos \theta \Bigg[\frac{2 M r \varpi^4+\Delta \Sigma^2}{\sqrt{\Lambda }
   \Sigma^2}\nonumber \\
&& -\frac{2 a^3  M r   \sin ^2\theta}{\Lambda (1 +\sqrt{Q} )} \sqrt{\frac{\Delta}{\Sigma^3}}\hat{P}_\phi\Bigg]\,,
\end{eqnarray}
\end{subequations}
\begin{subequations}
\begin{eqnarray}
F_t^2 &=& \frac{M \rho^2}{\Sigma^2}
    \Bigg[\frac{\varpi^2 \sqrt{\Sigma}}{\Lambda  
   \left(1+\sqrt{Q}\right) \sin\theta}\hat{P}_\phi - a \sqrt{\frac{\Delta}{\Lambda }}{\sin \theta} \Bigg]\,,\nonumber \\ \, \\
F_r^2 &=& \frac{a M  \left(2r^2\Sigma + \rho^2 \varpi^2\right) \sin \theta}{\Lambda 
   \left(1+\sqrt{Q}\right) \Sigma^{3/2}} \hat{P}_r \,,\\
F_\theta^2 &=& -\frac{2 a^3 M  r \Delta   \cos \theta \sin ^2\theta}{\Lambda (1 + \sqrt{Q} ) \Sigma^{3/2}}\hat{P}_r \,, \\
F_\phi^2 &=& -{\sin \theta} \Bigg[\frac{a M  (2r^2\Sigma + \varpi^2\rho^2)}{\Lambda  \left(1+\sqrt{Q}\right)\Sigma^{3/2}}\hat{P}_\phi \nonumber\\
&& +\sqrt{\frac{\Delta}{\Lambda }} \left(\frac{r \Sigma^2 - a^2 M \rho^2 \sin^2\theta}{\Sigma^2}\right)\Bigg]\,,
\end{eqnarray}
\end{subequations}
\begin{subequations}
\begin{eqnarray}
F_t^3 &=& -\frac{M}{\sqrt{\Lambda}(1+\sqrt{Q})\Sigma^{5/2}}\times\nonumber \\
&&\left(\rho^2 \varpi^2 \, \hat{P}_\theta  + 2a^2 r \Delta \sin\theta \cos\theta \, \hat{P}_r\right)\,,\\
F_r^3 &=& \frac{a^2 \cos \theta \sin \theta}{\sqrt{\Delta} \Sigma}\,,\\
F_\theta^3 &=& \frac{\sqrt{\Delta} r}{\Sigma}\,, \\
F_\phi^3 &=& \frac{a M \sin^2\theta}{\sqrt{\Lambda } \left(1+\sqrt{Q}\right) \Sigma^{5/2}} \Big[2 a^2 r \Delta \cos\theta \sin\theta  \hat{P}_r \nonumber \\
&& + (2r^2\Sigma + \rho^2 \varpi^2) \hat{P}_\theta \Big]\,.
\end{eqnarray}
\end{subequations}
Inserting these results into the Hamiltonian \eqref{Hamiltonian_canonical_final2}, a long but straightforward 
computation yields
\begin{equation}
\bar{H} = \bar{H}_{\rm NS} + \bar{H}_I S^I\,,
\end{equation}
where
\begin{widetext}
\begin{eqnarray}
\bar{H}_1 &=& -\left[\frac{\sqrt{\Delta}\cos\theta}{\Lambda^2\sqrt{\Sigma}\sqrt{Q}(1+\sqrt{Q})\sin^2\theta}\right]\Big[(1+\sqrt{Q})(\Delta\Sigma^2+2 M r \varpi^4) + 2a^2Mr\varpi^2 \sqrt{Q}\sin^2\theta\Big]\hat{P}_\phi \nonumber
\\
&& + \left[\frac{aM\Delta(2r^2\Sigma + \varpi^2\rho^2)\sin\theta}{\Lambda^{3/2}\Sigma^2\sqrt{Q}(1+\sqrt{Q})}\right]\hat{P}_r\hat{P}_\theta + \left[\frac{2a^3 M r \Delta \cos\theta \sin^2\theta}{\Lambda^{3/2} \Sigma \sqrt{Q}(1+\sqrt{Q})}\right]\left(1+\sqrt{Q} + \frac{2\Sigma}{\Lambda \sin^2\theta}\hat{P}_\phi^2 + \frac{\Delta}{\Sigma}\hat{P}_r^2 \right)\,,\\\nonumber\\
\bar{H}_2 &=& \left[\frac{\Delta(1+\sqrt{Q})(r\Sigma^2 - a^2M\rho^2\sin^2\theta) - M\sqrt{Q}(\rho^2\varpi^4 - 4a^2Mr^3\sin^2\theta)}{\Lambda^2\sqrt{\Sigma}\sqrt{Q}(1+\sqrt{Q})\sin\theta}\right]\hat{P}_\phi + \left[\frac{2a^3Mr\Delta^{3/2}\cos\theta\sin^2\theta}{\Lambda^{3/2}\Sigma^2 \sqrt{Q}(1+\sqrt{Q})}\right]\hat{P}_r\hat{P}_\theta \nonumber \\
&& + \left[\frac{aM\sqrt{\Delta}(2r^2\Sigma + \varpi^2\rho^2)\sin\theta}{\Lambda^{3/2}\Sigma \sqrt{Q}(1+ \sqrt{Q})}\right]\left(1 + \sqrt{Q} + \frac{2\Sigma}{\Lambda \sin^2\theta}\hat{P}_\phi^2 + \frac{1}{\Sigma}\hat{P}_\theta^2\right)\,, \\
\bar{H}_3 &=&  - \left[\frac{a^2 \Delta \cos\theta\sin\theta}{(\Lambda\Sigma)^{3/2}\sqrt{Q}(1+\sqrt{Q})}\right]\Big(\Lambda + \sqrt{Q}\Delta \Sigma \Big)\hat{P}_r - \left[\frac{r\Lambda\Delta + \varpi^2\Sigma \sqrt{Q}\big(r\Delta - M(r^2-a^2)\big)}{(\Lambda\Sigma)^{3/2}\sqrt{Q}(1+\sqrt{Q})}\right]\hat{P}_\theta \nonumber \\
&& -\left[\frac{aM\sqrt{\Delta}}{\Lambda^2 \Sigma \sqrt{Q}(1+\sqrt{Q})}\right]\Big[2a^2r\Delta \cos\theta\sin\theta \hat{P}_r + (2r^2\Sigma + \varpi^2\rho^2)\hat{P}_\theta\Big]\hat{P}_\phi \,.
\end{eqnarray}
\end{widetext}
Setting $a=0$ in this result and noting that for $a=0$ one has $\Lambda=r^4$, $\Sigma=r^2$ and $\Delta=r(r-2 M)$, it is easy to check that
this Hamiltonian reduces to the Schwarzschild result \eqref{H:schwarzschild} in the non-spinning case.

\section{Comparing the Hamiltonian in the generalized Newton-Wigner SSC with the ADM canonical Hamiltonian of 
PN theory}
\label{sec:hamiltonianPN}

In this section we specialize our results to the case of the Kerr spacetime, but this time using  ADM-TT coordinates. By expanding our Hamiltonian~\eqref{Hamiltonian_canonical_final} following the prescription of PN theory, we verify explicitly that we recover the
 known test-particle limit results of the Arnowitt-Deser-Misner (ADM) canonical 
 Hamiltonian  computed within PN theory alone. The latter is currently known through 2.5PN order for the terms linear in the 
 spin~\cite{Damour:2007nc}, and through 3PN order for the terms quadratic in the
 spin~\cite{PR06,PR07,SHS07,PR08b,ML08,SSH08,SHS08}. We cannot reproduce 
 the PN couplings of the test particle's spin with itself because the MPP equations,
 as we have already stressed, are only valid to linear order in the
 particle's spin. In addition we also obtain 
  the terms linear in the spins at 3.5PN order of the canonical ADM Hamiltonian in the test-particle limit. 
 Those contributions have never been  computed before.

In order to make the PN expansion as clear as possible, we restore factors of $c$ in this section. However these factors of $c$ must be viewed purely as dimensionless PN book-keeping parameters, and as such we are still formally employing geometric units.

First, let us introduce the Kerr metric in ADM-TT coordinates~\cite{Hergt:2007ha},
\begin{equation}\label{downmetric}
g_{\mu\nu}=
\begin{pmatrix}
-\alpha^2+\beta_{i}\beta^{i} & -\beta_{i}\\
-\beta_{j} & \gamma_{ij}
\end{pmatrix}\,,
\end{equation}
\begin{equation}
g^{\mu\nu}=
\begin{pmatrix}
-1/\alpha^{2} & -\beta^{i}/\alpha^2\\
-\beta^{j}/\alpha^2 & \gamma^{ij}-\frac{\beta^{i}\beta^{j}}{\alpha^2}
\end{pmatrix}\,,
\end{equation}
where $\gamma^{ik}\,\gamma_{kj}=\delta^i_j$ and $\beta^i=\gamma^{ik}\, \beta_k$.
Defining ${n}^i\equiv {x}^i/r$ and introducing a dimensionless three-vector 
$\boldsymbol{\chi}$ defined as
\begin{equation}\label{spin_par}
\boldsymbol{\chi}\equiv \frac{\boldsymbol{S}_{\rm Kerr}}{M^2}\,,
\end{equation}
where $M$ is the mass of the Kerr black hole and $\boldsymbol{S}_{\rm Kerr}$ its spin, the lapse function is given by~\cite{Hergt:2007ha}
\begin{eqnarray}
\alpha &=& c-\frac{M}{rc}+\frac{1}{2}\frac{M^2}{r^2c^3}-\frac{1}{4}\frac{M^3}{r^3c^5} \nonumber \\
&& + \frac{1}{8}\frac{M^4}{r^4c^7} + \frac{1}{2}\frac{M^3[3(\boldsymbol{\chi}\!\cdot\!\boldsymbol{n})^2-\chi^2]}{r^3 c^5} \nonumber \\
&& +\frac{1}{2}\frac{M^4[5\chi^2-9(\boldsymbol{\chi}\!\cdot\!\boldsymbol{n})^2]}{r^4c^7}+{\cal O}\left(9\right),\label{laadm}
\end{eqnarray}
the shift vector is given by
\begin{eqnarray}
\beta^{i}&=& \Bigg\{\frac{2M^2}{r^2 c^3}-\frac{6M^3}{r^3 c^5}+\frac{21}{2}\frac{M^4}{r^4 c^7} \nonumber \\
&& -\frac{M^4 [5(\boldsymbol{\chi}\!\cdot\!\boldsymbol{n})^2-\chi^2]}{r^4 c^7}\Bigg\}\epsilon^{ijk}\chi_{j}n_{k}+{\cal O}\left(9\right),\label{shadm}
\end{eqnarray}
and the spatial metric $\gamma^{ij}$ is given by
\begin{equation}\label{gammametric}
\gamma^{ij}= \frac{1}{A}\delta^i_j-\delta^{ik} \delta^{jl} h_{kl}^{\rm TT}+{\cal O}\left(10\right)\,,
\end{equation}
where $\epsilon_{ijk}=\epsilon^{ijk}$ is the Levi-Civita symbol (with
$\epsilon_{123}=\epsilon^{123}=1$), and where the quantities $A$ and $h_{kl}^{\rm TT}$ are defined as
\begin{eqnarray}
A &=& \left(1+\frac{M}{2r c^2}\right)^4+\frac{M^3[\chi^2-3(\boldsymbol{\chi}\!\cdot\!\boldsymbol{n})^2]}{r^3 c^6}  \nonumber\\
&& +\frac{1}{2}\frac{M^4\chi^2}{r^4c^8}-\frac{3M^4(\boldsymbol{\chi}\!\cdot\!\boldsymbol{n})^2}{r^4 c^8}\,,
\\
 h_{ij}^{\rm TT} &=& -\frac{7}{2}\frac{M^4\chi^2}{r^4c^8}\delta_{ij}+7\frac{M^4(\boldsymbol{\chi}\!\cdot\!\boldsymbol{n})^2}{r^4c^8}\delta_{ij} \nonumber \\
&& + 7\frac{M^4\chi^2n_{i}n_{j}}{r^4c^8}-21\frac{M^4(\boldsymbol{\chi}\!\cdot\!\boldsymbol{n})^2n_{i}n_{j}}{r^4c^8} \nonumber\\
&& +\frac{7}{2}\frac{M^4\chi_{i}\chi_{j}}{r^4 c^8}\,.
\end{eqnarray}
For the reference tetrad appearing in
the Hamiltonian, we chose
\begin{subequations}
\begin{eqnarray}\label{tetrad1}
\tilde{e}^{T}_{\mu} &=& \delta^t_\mu \alpha\,,\\\label{tetrad2}
\tilde{e}_{I}^{\mu} &=& \frac{\delta_I^\mu}{\sqrt{A}}+{\cal O}\left(8\right)\,.
\end{eqnarray}
\end{subequations}
It turns out however that we only need the spatial triad $\boldsymbol{\tilde{e}}_{I}$ through order $1/c^7$ for our purposes. (This makes the spatial triad very simple 
because the spatial metric is diagonal at that order).  

The canonical spin $S^I$ appearing in the Hamiltonian~\eqref{Hamiltonian_canonical_final} 
scales as the physical spin of the test particle. To conform with standard power counting
in PN theory, this spin variable carries a power of $1/c$. Therefore when restoring 
the factors of $1/c$ for the purpose of PN bookkeeping, we make the replacement
~\footnote{This is appropriate if
  the particle is a black hole or a rapidly rotating compact star. In
  the black hole case, $S=a m^2/c$, with $a$ ranging from $0$ to $1$
  [see Eq.~\eqref{spin_par}]. In the rapidly spinning star case one has $S=m v_{\rm rot} R\sim
  m c R_{s}\sim  m^2/c$ (where we have assumed that the rotational
  velocity $v_{\rm rot}$ is comparable to $c$ and that the stellar
  radius $R$ is of order of the Schwarzschild radius
  $R_s=m/c^2$).}
\begin{equation}
S^I \rightarrow \frac{S^I}{c}\,.
\end{equation}
Finally we define the orbital angular momentum as
\begin{equation}
L^i\equiv \epsilon^{ijk}\, x^j\, P_k\,,
\end{equation}
and rescaled momentum and spin as
\begin{subequations}
\begin{eqnarray}
\hat{\boldsymbol{P}} &=& \frac{1}{m}\boldsymbol{P}\,, \\
{\boldsymbol{S}}^\ast &=& \frac{M}{m}\boldsymbol{S}\,,
\end{eqnarray}
\end{subequations}
which are useful to abbreviate formulas below. With these tools it is straightfroward to expand the 
Hamiltonian~\eqref{Hamiltonian_canonical_final} in powers of $1/c$ as 
\begin{eqnarray}
\bar{H}&=&m\, c^2+ \bar{H}_{\rm N} +\frac{1}{c^2}\, \bar{H}_{\rm 1PN} +\frac{1}{c^3}\, \bar{H}_{\rm 1.5PN}+\frac{1}{c^4}\, 
\bar{H}_{\rm 2PN}\nonumber \\
&& +\frac{1}{c^5}\, \bar{H}_{\rm 2.5PN}+\frac{1}{c^6}\, \bar{H}_{\rm 3PN}
+\frac{1}{c^7}\, \bar{H}_{\rm 3.5PN} +{\cal O}\left(8\right)\nonumber \\
&& + {\cal O}(S^2) \,,
\label{Hamilt}
\end{eqnarray}
where
\begin{eqnarray}
\bar{H}_{\rm N} &=& m\left(\frac{\hat{\boldsymbol{P}}^2}{2} -\frac{M}{r}\right)\,, \\
\bar{H}_{\rm 1PN} &=& m\left(-\frac{\hat{\boldsymbol{P}}^4}{8}- \frac{3M}{2r} \hat{\boldsymbol{P}}^2 + \frac{M^2}{2 r^2}\right)\,, \\
\bar{H}_{\rm 1.5PN} &=& \frac{1}{r^3}\left(2\boldsymbol{S}_{\rm Kerr}+ \frac{3}{2}{\boldsymbol{S}}^\ast \right)\cdot \boldsymbol{L}\,,
\end{eqnarray}
\begin{eqnarray}
\bar{H}_{\rm 2PN} &=& m\Bigg(\frac{\hat{\boldsymbol{P}}^6}{16}+\frac{5M }{8r}\hat{\boldsymbol{P}}^4 +\frac{5M^2 }{2r^2}\hat{\boldsymbol{P}}^2 -\frac{M^3}{4 r^3}\Bigg) \nonumber \\
&& + \frac{m}{2Mr^3}(3n_{ij} - \delta_{ij})S^i_{\rm Kerr}\left(S^j_{\rm Kerr} + 2{S}^\ast_j 
\right) \,, \nonumber \\ \, \\
\bar{H}_{\rm 2.5PN} &=& \frac{1}{r^3}\left[-\frac{M}{r} \Big(6\boldsymbol{S}_{\rm Kerr} +5 \boldsymbol{S}^\ast\Big)  - \frac{5}{8}\hat{\boldsymbol{P}}^2 \boldsymbol{S}^\ast \right]\cdot\boldsymbol{L}\,, \nonumber \\
\end{eqnarray}
\begin{eqnarray}
\bar{H}_{\rm 3PN} &=& m\Bigg(-\frac{5\hat{\boldsymbol{P}}^8}{128}-\frac{7M}{16 r}\hat{\boldsymbol{P}}^6-\frac{27 M^2}{16r^2}\hat{\boldsymbol{P}}^4 \nonumber \\
&& -\frac{25M^3}{8r^3}\hat{\boldsymbol{P}}^2+\frac{M^4}{8 r^4}\Bigg)  + \frac{m}{2Mr^3}S^{ij}_{\rm Kerr} \times \nonumber \\
&& \left[\frac{3}{2}\hat{\boldsymbol{P}}^2\Big(3n_{ij} - \delta_{ij}\Big) - \frac{M}{r}\Big(9n_{ij} - 5\delta_{ij}\Big)\right] \nonumber \\
&& +\frac{3m n_{ij}}{2M r^3}\bigg[2\hat{{P}}^i{S}_{\rm Kerr}^k\, \hat{P}^{[j}S^{\ast\,k]} \nonumber \\
&&  - (\hat{\boldsymbol{P}}\times\boldsymbol{S}^\ast)^i(\hat{\boldsymbol{P}}\times\boldsymbol{S}_{\rm Kerr})^j
\bigg]\nonumber \\
&& +\frac{6 m}{r^4}S^{\ast\,i}S_{\rm Kerr}^j\left(\delta_{ij} - 2n_{ij}\right)\,, 
\end{eqnarray}
where $n_{ij} = n_in_j$ and $S^{ij}_{\rm Kerr} = S^i_{\rm Kerr} S^j_{\rm Kerr}$. The non-spinning terms in the Hamiltonian \eqref{Hamilt} coincide with the corresponding terms computed 
in PN theory in the test-particle limit~\cite{DJS3PN}; the linear terms in the spins at 1.5PN and 2.5PN order agree 
with the terms computed in the test-particle limit in PN theory~\cite{Damour-Schafer:1988,Damour:2007nc};  
the terms quadratic in the spin of the larger body coincide with what derived 
in PN theory at 2PN~\cite{Damour01c} and 3PN order~\cite{SSH08,SHS08}. We find that the contributions at 3.5PN are given by
\begin{eqnarray}
  \bar{H}_{\rm 3.5PN} &=& \frac{9m}{2 M^2 r^4}(\boldsymbol{S}_{\rm Kerr}\cdot\boldsymbol{n})(\boldsymbol{S}^\ast \times \boldsymbol{S}_{\rm Kerr})\cdot\hat{\boldsymbol{P}} \nonumber \\
  && -\frac{1}{4M^2r^5}\left[5(\boldsymbol{S}_{\rm Kerr} \cdot \boldsymbol{n})^2 -\boldsymbol{S}_{\rm Kerr}^2\right]
\left (9\boldsymbol{S}^\ast + \right. \nonumber \\
&& \left . 4\boldsymbol{S}_{\rm Kerr} \right )\cdot\boldsymbol{L} 
+ \frac{21M^2}{2r^5} \boldsymbol{S}_{\rm Kerr}\cdot\boldsymbol{L} \nonumber \\
  && + \Big (\frac{7}{16r^3}\hat{\boldsymbol{P}}^4 + \frac{27M}{8r^4}\hat{\boldsymbol{P}}^2 + \frac{75}{8} \frac{M^2}{r^5} \Big )
(\boldsymbol{S}^\ast\cdot\boldsymbol{L})\,. \nonumber \\ \label{H3.5PN}
\end{eqnarray}
While the terms of this expression which are cubic in the spins
(${S}_{\rm Kerr}^3$ and ${S}_{\rm Kerr}^2 {S}^\ast$) have already been
calculated for generic mass-ratios in Refs.~\cite{Hergt:2007ha,hergt_schafer_08}, with which we agree in the
test-particle limit, the terms linear in the spins are, as far as we
are aware, a new result. Of course, because our Hamiltonian is only valid at linear order in the particle's spin,
this result is still incomplete as it does not include terms $({S}^\ast)^3$ and ${S}_{\rm Kerr} ({S}^\ast)^2$, which
are still unknown.

Finally, we stress that at leading order our generalized NW SSC reduces to the
so-called baryonic SSC of Refs.~\cite{BOC79,Damour-Schafer:1988}. In fact, at leading order
$p_i\approx m v^i$, $p_t\approx -m c^2$ and $\tilde{e}^{T}_\mu\approx
c \delta^t_\mu$, which yields $\omega_t \approx -2 m c^2$ and
$\omega_i \approx m v^i$. Therefore, our generalized NW SSC becomes
\begin{equation}
S^{it}\approx \frac12 S^{ij}\frac{v^j}{c^2}\,,
\end{equation}
in agreement with Refs.~\cite{BOC79,Damour-Schafer:1988}.

\section{Conclusions}
\label{sec:conclusions}

In summary: starting from the Lagrangian put forward in Ref.~\cite{porto} building on the classical work of
Ref.~\cite{hanson} on the relativistic spherical top dynamics, we derived the 
unconstrained Hamiltonian for a spinning test-particle in a curved
spacetime, at linear order in the particle's spin. The equations of
motion for this Hamiltonian coincide with the MPP equations of motion.The latter are 
well-known to describe the motion and spin-precession of a test-particle, but are expressed in terms of the spin tensor $S^{\mu\nu}$ 
carrying six degrees of freedom. In order to eliminate three of these
degrees of freedom (which can be shown to correspond to the choice of
the point internal to the spinning body whose worldline is
followed~\cite{semerak}), we impose the so-called NW spin
supplementary condition, suitably generalized to curved spacetime. 
Using the formalism of Dirac brackets~\cite{Henneaux:1992ig} 
we computed the Hamiltonian and phase-space algebra of the constrained system. In particular, we
showed that, in a generic curved spacetime, the resulting phase-space algebra is canonical, 
\textit{i.e.} it has the standard sympletic structure for the set 
of dynamical variables $(\bm{q},\bm{p},\bm{S})$, at least at linear order in the particle's spin. 
As a consequence, the equations of motion can be derived from our constrained Hamiltonian by means of the usual well-known
Hamilton equations. 

As an application, making specific choices of the tetrad field,  
we computed explicitly the constrained 
Hamiltonian for a spherically symmetric spacetime, both in isotropic
and in spherical coordinates, as well as for the Kerr spacetime in
Boyer-Lindquist coordinates. We notice that different choices of the tetrad 
field would lead to different Hamiltonians connected by canonical transformations.
Also, we expanded our Hamiltonian in PN orders and showed explicitly that it reduces to the test
particle limit of the ADM canonical Hamiltonian  computed in PN
theory~\cite{Damour-Schafer:1988,Damour01c,Damour:2007nc,SSH08,SHS08}. 
Notably, we recover the known spin-orbit couplings through
2.5PN order and the spin-spin couplings of type $S_{\rm Kerr}\,S$
through 3PN order, $S_{\rm Kerr}$ being the spin of the Kerr
spacetime.  Our method allows one to compute the PN Hamiltonian, in
the test particle limit and at linear order in the particle's spin, at
\textit{any} PN order, and as an application we computed it at 3.5PN
order.

Another application of this work will be developed in a follow-up paper, 
where we will use our Hamiltonian to build a new effective-one-body 
Hamiltonian for spinning bodies~\cite{Buonanno99,DJS3PN,Damour01c,DJS08}. 
Such work will be important to build templates 
for the search of gravitational waves with ground and space-based detectors, 
as it will permit taking full advantage of the analytical and numerical 
treatment of the dynamics of spinning black-hole binaries throughout the 
inspiral, merger and ringdown phases.

\begin{acknowledgments}
  E.B., A.B. and E.R. acknowledge support from NSF Grant No. PHY-0603762. We would like to thank Ted Jacobson, Rafael
Porto and Jan Steinhoff for discussions, and Gerhard Schafer for useful comments.
  \end{acknowledgments}

\bibliography{references}

\end{document}